\documentstyle[preprint,aps,psfig]{revtex}
\newcommand{\beq}{\begin{equation}}
\newcommand{\eeq}{\end{equation}}
\newcommand{\beqar}{\begin{eqnarray}}
\newcommand{\eeqar}{\end{eqnarray}}
\tighten
\begin{document}
\draft
\preprint{Preprint Numbers: \parbox[t]{45mm}{KSUCNR-02-96\\
	                                  TRI-PP-96-25\\  nucl-th/9607025}}
\title{Pion Loop Contribution to \mbox{$\rho-\omega$} Mixing
						       and Mass Splitting}
\author{K. L. Mitchell\cite{Add} and P. C. Tandy}
\address{
Center for Nuclear Research, Department of Physics, Kent State University, 
Kent, Ohio 44242}

\date{July 12, 1996}
\maketitle
\begin{abstract}
We study the self-energy amplitudes for the $\rho-\omega$ meson
system produced by an effective field theory model at the quark level
characterized by a finite range quark-quark interaction and an isospin
breaking bare quark mass difference $m_u-m_d$.  Dynamical effects associated
with confinement and dressing of quarks and the finite size of the produced
$\bar{q}q$ meson modes are treated.
Parameters of this approach, previously constrained by
soft pion physics, are here supplemented in a minimal way to reproduce
$m_\omega$ and $g_{\rho\pi\pi}$.  Calculations are carried out up to the 
one pion loop level and the predictions for
$\rho^0-\omega$ mixing, mass splitting, the $\rho$ width and the
symmetry breaking $\omega\pi\pi$ coupling constant and form factor are
presented.  The contribution of the direct
\mbox{$\omega\rightarrow\pi\pi$} process relative to the
\mbox{$\omega\rightarrow\rho\rightarrow\pi\pi$} process is discussed within 
this model.

\end{abstract}
%
\section{Introduction}
\label{sec1}

Recent work on the pion and kaon systems has shown~\cite{pionff,Kem96} that a 
very efficient
description of masses, decay constants and electromagnetic form factors
in terms of QCD degrees of freedom can be implemented through a parameterized
form of the dressed quark propagator used in conjunction with 
truncations of the coupled Dyson-Schwinger equations (DSE).   Chiral symmetry   
considerations for these systems
eliminate the need for explicit information about gluon propagators and
the gluon-quark vertex  apart from that which is implicitly provided by the
dressed quark propagator.   At the simplest level this DSE approach~\cite{DSEr}
to the QCD modeling of hadron physics
has been successful in a number of applications involving the Goldstone bosons
including $\pi\pi$ scattering~\cite{pipi} and the anomalous 
$\gamma^* \pi \gamma$~\cite{FMRT95}, $\gamma \pi \pi \pi$~\cite{AR96}
and $\gamma \pi \rho$~\cite{M95,T96} processes.  The 
dynamical format of these initial studies can also be 
generated from a model field theory with QCD degrees of freedom containing
an effective finite range gluon propagator.  In particular, a treatment at
the $0^{th}$ order in the loop expansion of the produced $\bar{q}q$ Goldstone
meson modes will yield the same dynamical structure with a consistently 
specified content for the various dressed elements.   We take this latter point 
of view for the present work and consider a minimal extension beyond the 
Goldstone boson sector to address some interesting issues in the $\rho-\omega$ 
vector meson system.  A previous work on $\rho-\omega$ mixing with this 
approach,~\cite{MTRC} limited to $0^{th}$ order 
in meson loops, is here extended to the one pion loop level to address a wider
set of phenomena.   

Quark confinement should be an important element in the bound state dynamics 
and hadronic decays of the $\rho-\omega$ system.  Otherwise, with typical 
constituent $u$ and $d$ quark masses of about $350$~MeV, the nearby spurious 
$\bar{q}q$ production threshold would have undue influence.   The approach we  
take incorporates quark confinement through the absence of a mass-shell
pole in the employed dressed quark propagator.  The necessary momentum
dependence of the quark self-energy amplitudes implies a finite range for
the effective gluon propagator.  The associated quark momentum dependence 
of the meson-quark vertex functions reflects the finite size of the 
$\bar{q}q$ correlations.   To guide our development of an effective action 
for $\pi$, $\rho$ and $\omega$ that preserves these realistic features 
we use the two flavor version of the Global Color Model (GCM)~\cite{CR85,GCM}. 
The employed dressed quark propagator model is constrained by soft chiral 
physics and is a simplified version of what is currently 
available~\cite{pionff,Kem96} in order
to facilitate the type of calculations necessary for this work.
Information about the dressed gluon propagator, necessary to treat
mesons that are not Goldstone mesons, is here introduced in a minimal way 
that is constrained by calculated values of $m_\omega$ and $g_{\rho\pi\pi}$. 

The zero range limit of the model we employ is very closely related to the
Nambu--Jona-Lasinio model (NJL)~\cite{NJL} which has been extensively used
to model hadron physics.     
However, in contrast with the present approach based upon the GCM, 
many important loop integrals encountered within the NJL model must be  
regulated via additional parameters.  Also quark confinement cannot be 
generated from the contact NJL interaction and the dynamically generated quark 
mass is a constant (constituent mass) rather than the (running) mass function
which is a necessary consequence of finite range gluon propagation.  The 
associated nonlocalities have a profound effect upon the nature of the 
calculations and the role that is assigned to phenomenology.    
A previous study of the $\rho-\omega$ system within the NJL model was limited
to the quark loop level for the mixing amplitude, but did generate the $\rho$ 
propagator, including the $2\pi$ width, at the pion loop level~\cite{FR95}.  
In Refs.~\cite{GSW96,SS96} an extended NJL model that includes an extra 
confining interaction has been applied to the $\rho-\omega$ system in a way 
that is the most comparable to the approach we take.

One of the topics we investigate is $\rho^0-\omega$ mixing generated by the 
bare quark mass difference $m_d-m_u$.   Recent interest in a description of 
$\rho^0-\omega$ mixing from the quark level 
has centered around the momentum dependence of the mixed self-energy amplitude
and the possible contribution to the charge symmetry breaking (CSB) component
of the nucleon-nucleon force.\cite{MTRC,GHT,Krein}  The
quark loop mechanism produces a momentum dependence that suppresses the 
amplitude at the small space-like momenta appropriate to the mixed 
$\rho-\omega$ exchange mechanism.  Here we investigate the one pion loop 
correction and find that the space-like mixing still remains negligibly
small.  A more promising explanation for
the CSB nuclear force has been found in the meson-nucleon vertex~\cite{GHP95}.

Our study of the pion loop contributions to the self-energies of the 
$\rho-\omega$ system produces calculations of the $\rho\pi\pi$ and 
$\omega\pi\pi$ coupling constants and form factors as well as the $2\pi$
width of the $\rho$ and the mass splitting $m_\rho-m_\omega$.  Initial
studies within the GCM of $m_\rho-m_\omega$~\cite{msplit}  and 
$g_{\rho\pi\pi}$~\cite{rpp87}  approximated hadron mass-shell information 
by behavior at zero meson momenta.  This
was to avoid ambiguities associated with numerical extrapolations of dressed 
quark propagators to the complex momenta needed to put hadron momenta 
on-shell in this Euclidean space approach.  The correct mass-shell 
definitions can be maintained in the present work due to the assured analytic
properties of the model quark propagator employed.

The amplitude for the direct process $\omega_I \rightarrow \pi~\pi$ is 
conventionally assumed~\cite{R}
to cancel out in the analysis of $e^+~e^- \rightarrow \pi~\pi$ data to
extract a value for the physical $\rho^0-\omega$ mixing amplitude~\cite{coon}.  
Here $\omega_I$ is the pure isospin component of 
the physical $\rho$.  However, it has
recently been argued~\cite{MOW96} that, without a theoretical model for the 
$\omega_I \rightarrow \pi~\pi$ amplitude, the error on the extracted mixing 
amplitude is much greater than previously thought.  Evidence for a significant 
contribution from the $\omega_I\pi\pi$ process is  obtained in a recent QCD 
sum rule analysis of the mixed-isospin vector current correlator~\cite{M96}.
In contrast, the value for $g_{\omega_I\pi\pi}$ recently estimated from the 
extended NJL model of Ref.~\cite{SS96} suggests a negligible effect.  
In the present work, the strength of this 
G-parity violating $\omega_I\pi\pi$ coupling, as driven by $m_d-m_u$ in the 
underlying quark loop, is found to be significant enough
to require an increase of typically $20\%$ in the required mixing amplitude.

In Sec.~\ref{sec2} we describe the underlying quark model action and also 
general aspects of the resulting effective meson action that we shall use.
The connection between them, i.e. the generation of $\bar{q}q$ meson modes 
from the GCM action, has been covered in previous works and key elements are
here reviewed briefly in the Appendix.   
Specific aspects of the pion sector, and the model dressed quark
propagator which is constrained by it , are described in Sec.~\ref{sec3}.  The
mixed $\rho-\omega$ meson sector produced at tree level by integrating out the
quarks is described in detail in Sec.~\ref{sec4}.  At that same level, the 
produced $\rho\pi\pi$ and $\omega\pi\pi$ interactions are identified in 
Sec.~\ref{sec5} and the calculated coupling constants and form factors are
described. 
The pions are integrated out up to the one loop level in 
Sec.~\ref{sec6} and the contributions to the mixing as well as to the $\rho$ 
width and mass shift are described.  
Also presented in Sec.~\ref{sec6} is a discussion of
the relative contribution of $\rho^0-\omega$ mixing and the
direct process $\omega_I \rightarrow \pi~\pi$ to the physical coupling strength 
$g_{\omega\pi\pi}$.  Further discussion follows in Sec.~\ref{sec7}.

\section{Quark Model}
\label{sec2}

To obtain a practical covariant description of both meson
substructure and interactions in terms of dynamically dressed quark degrees of
freedom we use the Global Color Model~\cite{CR85,GCM}.   
This is an effective field theory at the quark level described by the action
\beq
S[\bar{q},q]=\int d^{4}x~\bar{q}(x) 
\bigl( \gamma \cdot \partial _{x} + m \bigr) q(x)
+ \frac{1}{2} \int d^{4}x d^{4}y~
	 j_{\mu }^a(x) D_{\mu\nu}(x-y)j_{\nu }^a(y)
\label{sgcm}
\eeq
where the quark currents, 
$j_\mu^a(x)=\bar{q}(x)\frac{\lambda^a}{2}\gamma_\mu q(x)$,
interact via a phenomenological gluon two-point function 
(\mbox{$ D_{\mu\nu}(x)=\delta_{\mu\nu} D(x)$} in the Feynman-like gauge
chosen here).    
We consider only two flavors ($u$ and $d$)
and use a Euclidean metric throughout, such that $a \cdot b= a_\mu b_\mu$ and
$\{\gamma_\mu ,\gamma_\nu \}=2\delta_{\mu\nu}$, with 
$\gamma_\mu= \gamma_\mu^\dagger$.  The bare masses are represented by
the diagonal matrix \mbox{$m=(m_u,m_d)$}.   
The closely related NJL model~\cite{NJL} is specified by the zero-range limit
of Eq.~(\ref{sgcm}) together with a set of coupling constants for the 
independent meson modes.  

There has been extensive work on both the meson and baryon sectors that arise
from hadronization of the GCM and a comprehensive review~\cite{rtc} is 
available.  For just the lowest mass modes,
the meson action for the effective local field variables 
\mbox{$\vec{\pi}, \vec{\rho}_\mu, \omega$} can be written as~\cite{CR85,GCM}
\beqar
\hat{S}[\pi,\rho,\omega]&=& {\rm Tr} \sum_{n=2}^\infty\frac{(-)^n}{n}
[S_0(i\gamma_5\vec{\tau}\cdot\vec{\pi}\hat{\Gamma}_\pi 
+i\gamma_\mu\omega_\mu \hat{\Gamma}_V 
+i\gamma_\mu\vec{\tau}\cdot\vec{\rho}_\mu \hat{\Gamma}_V)]^n \nonumber \\  
& &+ 9\int d^4x~d^4y \;
\frac{\frac{1}{2}\vec{\pi}\cdot\vec{\pi} \, \hat{\Gamma}_\pi^2    
+\omega^2 \, \hat{\Gamma}_V^2 +\vec{\rho}\cdot\vec{\rho} \, \hat{\Gamma}_V^2}
{2D(x-y)}.
\label{mesonaction}
\eeqar
For completeness, we collect the main steps in the transition from 
Eq.~(\ref{sgcm}) to Eq.~(\ref{mesonaction}) into the Appendix.  In 
Eq.~(\ref{mesonaction}) Tr denotes the trace over color, flavor and spin as 
well as the continuous space-time variable.  The quark propagator is 
\mbox{$ S_0(p) =$} \mbox{$[i\not \! p A(p^2) +B(p^2)+m]^{-1}$}.  
The first term of Eq.~(\ref{mesonaction}) represents the sum of all possible
meson couplings mediated by a single quark loop and the second term can be
described as a contribution to the meson mass terms.  

For simplicity, we have made the
common approximation in  Eq.~(\ref{mesonaction}) that each meson 
Bethe-Salpeter (BS) amplitude is truncated to its single canonical Dirac 
matrix covariant with $\hat\Gamma$ being the associated scalar function. In
studies of the BS equation that allow a more complete selection of 
covariants,\cite{JM93,sep96} this is found to be the dominant component.  
In Eq.~(\ref{mesonaction}) we have also used the result that for the pure
isospin components \mbox{$\hat{\Gamma}_\rho = \hat{\Gamma}_\omega $} 
which follows from the ladder BS equation when coupling terms
generated from $m_d- m_u$ are ignored.  This defines the vector meson states
of good isospin whose mixing we shall subsequently study.   In a momentum
representation the meson-quark coupling implicit in Eq.~(\ref{mesonaction}) is,
e.g., \mbox{$\bar{q}(q_+) i\gamma_\mu \omega_\mu(P)\hat{\Gamma}_V(q;P)q(q_-)$}
where the quark momenta are $q_\pm =q \pm P/2$ with $P$ being the meson
momentum.  Thus $\hat{\Gamma}_V(q;P)$ plays the role of a 
form factor but has, in principle, a well-defined dynamical
content defined by the model at the quark-gluon level.  In this 
work we utilize the dynamical structure of Eq.~(\ref{mesonaction}) to 
combine previous work on the pion sector with
a phenomenological approach to $\hat\Gamma_V$ to explore whether 
important features of the $\rho-\omega$ system can be described.

\section{The Pion Sector}
\label{sec3}

The free pion part of the action in Eq.~(\ref{mesonaction}) is
\beq
\hat{S}_2[\pi]=\frac{1}{2}\int\frac{d^4P}{(2\pi)^4}
\vec{\pi}(-P) \cdot \vec{\pi}(P) \; \hat{\Delta}^{-1}_\pi(P^2), 
\label{freepi}
\eeq
where the inverse propagator for the effective local pion field is
\beq
\hat{\Delta}^{-1}_{\pi}(P^2) = {\rm tr} \int \frac{d^4q}{(2\pi)^4}
[S_0(q_-)i\gamma_5\tau S_0(q_+)i\gamma_5\tau] \; 
\hat{\Gamma}_\pi^2(q;P)
+\frac{9}{2}\int d^4 r\frac{\hat{\Gamma}_\pi^2(r;P)}{D(r)}.
\label{delpi}
\eeq
Here \mbox{$q_\pm=q\pm\frac{P}{2}$}
and $\tau$ is any single component of $\vec{\tau}$.
Throughout this work the symbol tr denotes a trace over color, flavor 
($N_f=2$) and spin.  We note that $\hat{\Delta}_\pi^{-1}$ is a 
functional of $\hat{\Gamma}_\pi$ and that 
\mbox{$\delta \hat{\Delta}_\pi^{-1}/\delta\hat\Gamma_\pi =0$}  is the
ladder BS equation (projected onto  the $i\gamma_5\tau$ channel) for the
amplitude $\hat{\Gamma}_\pi$.     To produce $\hat{\Gamma}_\pi$ that way
requires a specification of the model gluon propagator. 
However an indirect route is provided through chiral symmetry  by the
Goldstone theorem in terms of the dressed quark propagator.
The realization relevant to the present treatment is that,
in the chiral limit, the ladder BS equation for the amplitude 
$\hat\Gamma_\pi(q;P)$ becomes identical to the ladder Dyson-Schwinger equation 
for \mbox{$B_0(q^2)\equiv B(q^2;m=0)$}, the scalar part of the 
dynamically generated quark self-energy~\cite{DelScad}.  
Thus the chiral limit pion, in this approach, is automatically both a massless 
Goldstone boson and a finite size $q\bar{q}$ 
bound state.  For a small bare mass, numerical studies~\cite{FR96}
reveal that $\hat{\Gamma}_\pi$ is still well approximated by the scalar quark 
self-energy amplitude.   

In this work we take \mbox{$\hat{\Gamma}_\pi(q;P) \approx B(q^2;m)$} and  
use \mbox{$m_{av} = (m_u +m_d)/2$} as the bare mass for both the $u$ and $d$.
Near the mass shell, we have the general form 
\mbox{$\hat\Delta^{-1}_\pi(P^2) = (P^2+m_\pi^2)f_\pi^2$}, where $f_\pi$ is
the pion wavefunction renormalization constant.   The
(constant) second term of Eq.~(\ref{delpi}) may therefore be eliminated 
if we require that the GMOR relation~\cite{GMOR} in the form 
\mbox{$\Delta_{\pi}^{-1}(0) = 2m_{av} |\langle\bar{q}q\rangle| $} be
satisfied.  Here $\langle\bar{q}q\rangle$ is the isospin average 
vacuum condensate.  In this regard we note that recent 
investigations~\cite{FR96,CG95a} find that a somewhat differently defined 
quantity is better suited to this relation.  The numerical consequences for 
the present work are not significant.  Our procedure yields
\begin{equation}
\hat{\Delta}^{-1}_{\pi}(P^2) = 2m_{av} |\langle\bar{q}q\rangle| - {\rm tr}
\int \frac{d^4q}{(2\pi)^4} [S_0(q_-)S_0(-q_+) - S_0(q)S_0(-q)] \; B^2(q^2)  ,
\label{delpigmor}
\end{equation}
where we take \mbox{$|\langle\bar{q}q\rangle| = 
\frac{1}{2}~{\rm tr} \int^{\mu^2} \frac{d^4q}{(2\pi)^4} S_0(q)$} with the 
scale specified
by the cutoff $\mu^2 = 1~{\rm GeV}^2$.  Now  $\hat\Delta^{-1}_\pi$ is fully 
specified in terms of the quark propagator.
Numerical evaluation of Eq.~(\ref{delpigmor}) identifies the pion mass 
through the zero at \mbox{$P^2=-m_\pi^2$}.

We also identify $Z_\pi(P^2)$ defined by
\mbox{$\hat\Delta^{-1}_\pi(P^2) = (P^2+m_\pi^2)Z_\pi(P^2)$}.  The momentum 
dependence of $Z_\pi(P^2)$ implies a dynamical mass function $m_\pi(P^2)$
due to $\bar{q}q$ substructure.  
We calculate $f_\pi$ from
$f_\pi^2 = Z_\pi(P^2=-m_\pi^2)$.  A physical normalization (unit residue at the
mass-shell pole) is produced by absorbing $f_\pi$ into the pion fields. For 
convenience, we redefine the fields to absorb $\sqrt{Z_\pi(P^2)}$  
so that the propagator becomes $\Delta_\pi=(P^2+m_\pi^2)^{-1}$.  The 
associated vertex amplitude is then the dimensionless quantity
\mbox{$\Gamma_\pi(p;P)=B(p^2)/\sqrt{Z_\pi(P^2)}$}.   At the mass-shell this 
procedure satisfies the canonical Bethe-Salpeter normalization 
condition~\cite{IZ}.  For any pion momentum, this choice defines 
the effective pion -quark vertex and related pion field consistent with 
the point-pion  propagator $\Delta_\pi=(P^2+m_\pi^2)^{-1}$ to  
allow contact with common procedures in nuclear physics.  In various  
applications we test the role of the composite nature of the pion in 
by replacing the function $Z_\pi(P^2)$ by its 
mass-shell value $Z_\pi(-m_\pi^2)$.  Such tests will be described later. 

For the dressed quark propagator, we use the
representation \mbox{$S_0(p)=$} \mbox{$-i \gamma\cdot{p}~\sigma_{V}(p^2) +$}
\mbox{$\sigma_{S}(p^2)$}, and employ the following simplified form of 
previously developed parameterized amplitudes~\cite{pionff}
\beqar
\bar{\sigma}_{S}(x)&=&c(\bar{m})~e^{-2x}+\frac{\bar{m}}{x}
\left( 1-e^{-2x} \right) \label{sigmas},       \\
\bar{\sigma}_{V}(x)&=&\frac{ 2x-1 + e^{-2x} } {2x^2} 
- c(\bar{m})~\bar{m}~e^{-2 x} \label{sigmav},
\eeqar
with $x=p^2/\lambda^2$, $\bar{\sigma}_{S}=\lambda \sigma_{S}$,
$\bar{\sigma}_{V}=\lambda^2 \sigma_{V}$, $\bar{m}=m/\lambda$, where $m$ is
the bare quark mass and $\lambda$ is the momentum scale.  This $S_0(p)$ is an
entire function in the complex momentum plane, a condition sufficient to
ensure the absence of quark production thresholds in S-matrix amplitudes for
physical processes~\cite{RWK}.  The original parameterization is guided by a 
confining model DSE~\cite{BRW} and by the behavior found in realistic DSE 
studies~\cite{DSEr,WKR}.  It is consistent with pQCD in the deep
Euclidean region apart from $\ln(p^2)$ corrections.  The parameters are
$\lambda=0.889~{\rm GeV}$, 
$~c(\bar{m}_{av})=c_0=0.581$, $~m_{av} = 16~{\rm MeV}$.  The 
soft chiral physics quantities produced by these parameters are
$f_{\pi} = 90.1$~MeV, $m_{\pi} = 143$~MeV, and 
$\langle\bar{q}q\rangle=(173~{\rm MeV})^3$.  The more elaborate original 
parameterization~\cite{pionff} provides a better fit to the above and also 
an excellent description of the pion electromagnetic 
form factor and the $\pi\pi$ scattering lengths.  

With the isoscalar quark propagator set as above, 
the isovector term is generated from the bare mass
dependence by substitution of
$\bar{m}=\bar{m}_{av}+\frac{\tau_3}{2}\delta\bar{m}$ with 
$\delta\bar{m}=\bar{m}_u-\bar{m}_d$.  
The constant $c(\bar{m})$ mainly represents the strength of dynamical chiral 
symmetry breaking that generates the dynamical mass function, 
$M(p^2)=\sigma_S(p^2)/\sigma_V(p^2)$. However a weak 
bare mass dependence for $c(\bar{m})$ is necessary to simulate 
the bare mass dependence of realistic DSE solutions.
This leads to a small isovector component of $c(\bar{m})$ and we write 
\mbox{$c(\bar{m})= c_0 +\tau_3 c_1$}, where $c_0 = c(\bar{m}_{av})$ is given
above, and $c_1 =(c_u-c_d)/2$.  To determine the latter we fit to the bare mass
dependence of $\bar{\sigma}_S(p^2=0)$ produced by realistic DSE 
studies~\cite{WKR}.  Only a linear behavior $c_1=d \delta\bar{m}/2$ 
characterized by the slope $d = c'(\bar{m})<0$ is important, and with 
$\delta m = -4~{\rm MeV}$, we have $c_1=0.0048$.  This  $1\%$ 
isospin-breaking effect $c_1/c_0$ is of the order of magnitude expected.
The quark propagator can be written as
\mbox{$S_0(p)=S_0^0(p)+\tau_3 S_0^1(p)$} where the latter isovector component
is responsible for the various $\rho^0-\omega$ mixing mechanisms considered 
here.

\section{The $\rho-\omega$ Sector at Tree Level}
\label{sec4}

We treat the transverse modes in the $\rho$ and
$\omega$ channels and use fields that contain the transverse projector
\mbox{$T_{\mu \nu}(P)= \delta_{\mu \nu} -P_{\mu}P_{\nu}/P^2$}.   That is, 
$\vec{\rho}_\mu(P) \equiv T_{\mu\nu}(P)\vec{\rho}_\nu(P)$ and 
$\omega_\mu(P) \equiv T_{\mu\nu}(P)\omega_\nu(P)$.  
In a matrix notation where $V_{\mu}$ denotes  $(\vec{\rho}_{\mu}, 
\omega_{\mu})$,  the tree-level action from Eq.~(\ref{mesonaction}), 
up to second order in the fields, can be written as 
\beq
\hat{S}_2[\rho,\omega] = \frac{1}{2} \int \frac{d^4P}{(2\pi)^4}
V^T_{\mu}(-P) \Bigl[ \hat{\Delta}^{-1}_{\mu \nu}(P) 
+ \hat{\Pi}_{\mu \nu}^q(P) 
\Bigr] 
V_{\nu}(P) . 
\label{sromix}
\eeq
Here, in meson channel space, $\hat{\Delta}^{-1}$ is diagonal 
and the only non-zero elements of $\hat{\Pi}^q$ are the off-diagonal ones 
that provide $\rho^0 -\omega$ coupling.   The superscript $q$ identifies 
the quark-loop contribution to distinguish from a  
different contribution introduced later.
The diagonal inverse propagator is given by 
\beq
\hat{\Delta}^{-1}_{\mu \nu}(P) = {\rm tr} \, \int \frac{d^4q}{(2\pi)^4}
\hat{\Gamma}_V^2(q;P)
\, \left[ S_0(q_-)i\gamma_\mu f S_{0}(q_+)i\gamma_\nu f\right] 
+ 9\,\delta_{\mu \nu} \int d^4 r \, \frac{\hat{\Gamma}_V^2(r;P) }{D(r)}.
\label{ipropv}
\eeq
Here $f$ 
are the relevant flavor factors ($1_f$ for the $\omega$, and $\vec{\tau}$ for 
the $\vec{\rho}$) and $q_\pm=q\pm\frac{P}{2}$, where $P$ is the meson 
momentum and $q$ is the loop momentum, or equivalently the $\bar{q}q$ 
relative momentum.   The amplitude $\hat{\Gamma}_V$ is defined in the diagonal
sector by the ladder BS equation produced by
\mbox{$\delta \hat{\Delta}^{-1}/\delta \hat{\Gamma}_V = 0$}.  Thus 
$\hat{\Gamma}_V$ contains no isospin mixing mechanism and is the same for 
$\rho$ and $\omega$.  The diagonal $\rho$ and $\omega$ elements of the 
inverse propagator are likewise identical.   
The transverse component $\hat{\Delta}^{-1}(P^2)$, defined by 
\mbox{$\hat{\Delta}^{-1}_{\mu\nu}(P) = T_{\mu\nu}(P)\hat{\Delta}^{-1}(P^2)$},
is given explicitly by
\beqar
\hat{\Delta}^{-1}(P^2)&=& 24\int \frac{d^4q}{(2\pi)^4}
\hat{\Gamma}_V^2(q;P)\left\{\sigma_V(q_+^2)\sigma_V(q_-^2)
\left[ \frac{P^2}{4} + \frac{q^2}{3} - \frac{2(P\cdot q)^2}{3P^2} \right]
-\sigma_S(q_+^2)\sigma_S(q_-^2)
\right\} \nonumber \\
&+& 9 \int d^4 r \frac{ \hat{\Gamma}_V^2(r;P) }{D(r)} .
\label{delt}
\eeqar
Here we have ignored a small contribution from the isovector component of
the quark propagator by using $m_{av}$ as the bare quark mass. 

The mixed self-energy, $\hat{\Pi}_{\mu \nu}^q(P)$, is produced from the same 
quark loop term of the effective action of Eq.~(\ref{mesonaction}) that 
produces $\hat{\Delta}^{-1}_{\mu\nu}$.  The mechanism is illustrated in 
Fig.~1.  The explicit expression is
\beq
\hat{\Pi}_{\mu \nu}^q(P) = {\rm tr}\, \int \frac{d^4q}{(2\pi)^4}\,
\left[ S_{0}(q_{-})i\gamma_{\mu} \tau_{3} S_{0}(q_{+})i\gamma_{\nu}\right] 
\hat{\Gamma}^2_V(q;P),
\label{pimn}
\eeq
which is identical to the first term of Eq.~(\ref{ipropv}) except that one of 
the flavor factors is $\tau_{3}$ and the other is $1_f$.  Thus, 
$\hat{\Pi}_{\mu \nu}^q(P)$ is nonzero only if $S_{0}$ has an isovector 
component. 
The structure of Eq.~(\ref{pimn}) is corresponds to first-order perturbation 
theory for isospin breaking effects in that the internal meson states or BS 
amplitudes that enter are from the unperturbed or isospin symmetric sector.  
The mixed self-energy may be expressed as the difference of quark loops
and the leading contribution is proportional to $\delta m$.   
The transverse component is
\mbox{$\hat{\Pi}^q(P^2) = \hat{\Pi}_u - \hat{\Pi}_d$}, where
\beq
\hat{\Pi}_f(P^2)= 12\int \frac{d^4q}{(2\pi)^4}
\hat{\Gamma}_V^2(q;P)\left\{\sigma^f_V(q_+^2)\sigma^f_V(q_-^2)
\left[ \frac{P^2}{4} + \frac{q^2}{3} - \frac{2(P\cdot q)^2}{3P^2} \right]
-\sigma^f_S(q_+^2)\sigma^f_S(q_-^2)  \right\} .
\label{pit}
\eeq
To summarize the transverse $\rho^0-\omega$ sector, the inverse propagator 
obtained at the quark loop level may be written  as
\beq
\hat{{\cal D}}^{-1}(P^2) = \left( \begin{array}{cc} 
\hat{\Delta}^{-1}(P^2) & \hat{\Pi}^q(P^2) \\
\hat{\Pi}^q(P^2)       & \hat{\Delta}^{-1}(P^2)  \end{array}  \right) .
\label{qlinvp}
\eeq
We shall refer to the basis here as ($\rho^0_I,\, \omega_I$) to indicate 
pure isospin states. 

The composite $\bar{q}q$ nature of the mesons is 
reflected in the fact that the diagonal elements contain a dynamical
mass function.  The general form 
\beq
\hat{\Delta}^{-1}(P^2)\, = \,(P^2+m_V^2)Z_1(P^2)
\, = \, (P^2+ m_V^2(P^2)) 
\eeq
identifies the degenerate mass $m_V$ of the $\rho^0_I$ and $\omega_I$ 
states.  
It is convenient to carry out part of the necessary 
renormalization  at this stage.  Without the off-diagonal terms of 
Eq.~(\ref{qlinvp}),  a
physical normalization (unit residue at the mass-shell pole) would be 
produced by absorbing at least the on-mass-shell value of
$\sqrt{Z_1}$ into the fields.  As with the pion sector,
we choose to absorb the function $\sqrt{Z_1(P^2)}$ into the fields
so that the resulting unmixed propagators have the standard point meson form.  
to
The associated BS amplitudes are now 
\mbox{$\Gamma_V(q;P)= \hat{\Gamma}_V(q;P)/ \sqrt{Z_1(P^2)}$}, which are
dimensionless quantities that, at the mass-shell, satisfy the standard 
normalization condition~\cite{IZ} for a BS amplitude.  We note that the 
meson-$\bar{q}q$ coupling strength (magnitude of $\Gamma_V$ at the meson 
mass-shell) is thus a prediction of this approach.  
The quark loop contribution to the $\rho^0-\omega$ sector is now
\beq
{\cal D}^{-1}(P^2) = \left( \begin{array}{cc} 
P^2+m_V^2        & \Pi^q(P^2) \\
\Pi^q(P^2)       & P^2+m_V^2  \end{array}  \right)  ,
\label{qlinvpr}
\eeq
where \mbox{$\Pi^q(P^2) = \hat{\Pi}^q(P^2)/Z_1(P^2)$}.  
The role played by the dynamical $\bar{q}q$ 
substructure of propagators may be obtained by holding the function 
$Z_1(P^2)$ at its on-mass-shell value to simulate a structureless meson.  

To facilitate these exploratory calculations, we adopt the simple form
\mbox{$\hat{\Gamma}_V(q;P)=$} \mbox{$N~exp(-q^2/a^2)$} which has previously 
been found to provide an adequate one-parameter representation of the 
relevant BS 
amplitude in a variational approach~\cite{rtc}.   The physical amplitude  
$\Gamma_V$ is independent of $N$.  The choice of effective
gluon propagator $D(r)$ in Eq.~(\ref{delt}) is made indirectly by ensuring 
that the resulting constant value of the second term produces a zero in
$\hat{ \Delta}^{-1}(P^2)$ at \mbox{$P^2=-m_V^2$} where
\mbox{$m_V = m_\omega =782~{\rm MeV} $}.  The range parameter $a$ is chosen 
to reproduce the empirical value of $g_{\rho\pi\pi}$ as discussed in the next 
Section.

The transverse mixing amplitude  
$\Pi^q (P^2)$ obtained from this approach has been discussed in a previous 
work~\cite{MTRC}.   The present parameterization of the input to this model 
is slightly different but only small 
quantitative changes occur in the calculated $\Pi^q (P^2)$ and
the conclusions of our previous work remain unchanged.  

\section{The $\rho\pi\pi$ and $\omega\pi\pi$ Interactions}
\label{sec5}

{}From the effective action in Eq.~(\ref{mesonaction}), we identify the 
interaction term
\beq
S\left[\rho\pi\pi\right]=
-{\rm Tr}\left[S_0 \, i\gamma_\mu \vec{\tau}\cdot\vec{\rho}_\mu \Gamma_V
(S_0 \, i\gamma_5 \vec{\tau}\cdot\vec{\pi}\Gamma_{\pi})^2 \right] .
\eeq
Use of the momenta shown in the quark triangle diagram of
Fig.~\ref{rpp_tri} yields the explicit expression
\beq
S\left[\rho\pi\pi\right]= i \int \frac{d^4P,Q}{\left[(2\pi)^4\right]^2}
\, \vec{\rho}_\mu(Q)\cdot \vec{\pi}(-P-\frac{Q}{2})
\times\vec{\pi}(P-\frac{Q}{2})
\, \Lambda_\mu^\rho(P,Q) ,
\label{srpp}
\eeq
where the vertex is
\beq
\Lambda_\mu^\rho(P,Q)=\int \frac{d^4k}{(2\pi)^4}
\Gamma_V(k+\frac{P}{2};Q) \Gamma_\pi(k+\frac{Q}{4};-P-\frac{Q}{2})
\Gamma_\pi(k-\frac{Q}{4};P-\frac{Q}{2}) T_\mu(k,P,Q) 
\label{vrpp}
\eeq
with the discrete loop trace denoted by
\beq
T_\mu(k,P,Q) = {\rm tr}\left[
S_0(k+\frac{P}{2}+\frac{Q}{2}) \, i\gamma_\mu 
S_0(k+\frac{P}{2}-\frac{Q}{2}) \, i\gamma_5 
S_0(k-\frac{P}{2}) \, i\gamma_5 \right].
\eeq
The symmetry properties of $\Lambda_\mu^\rho(P,Q)$ that follow are  
$\Lambda_\mu^\rho(P,Q)=-\Lambda_\mu^\rho(-P,Q) =\Lambda_\mu^\rho(P,-Q)$.  
Hence the most general form is
\beq
\Lambda_\mu^\rho(P,Q)=-P_\mu \; F_{\rho\pi\pi}(P^2,Q^2,(P\cdot Q)^2)
-Q_\mu P\cdot Q \; H_{\rho\pi\pi}(P^2,Q^2,(P\cdot Q)^2).
\label{rppgen}
\eeq

With both pions on the mass-shell, 
\mbox{$(P-\frac{Q}{2})^2=(P+\frac{Q}{2})^2=-m_\pi^2$}.  Equivalently,  
$P\cdot Q=0$ and \mbox{$P^2=-m_\pi^2-\frac{Q^2}{4}$} so that the simplified 
form of Eq.~(\ref{rppgen}) is \mbox{$\Lambda_\mu(P,Q)
= -P_\mu F_{\rho\pi\pi}(Q^2)$}.  The 
coupling constant is defined as the mass-shell value
\mbox{$g_{\rho\pi\pi} = F_{\rho\pi\pi}(Q^2=-m_\rho^2) $}.  If the form factor is 
held at this value for all momenta, one obtains the point coupling limit 
\beq
S\left[\rho\pi\pi\right]= -ig_{\rho\pi\pi} \int
\frac{d^4P,Q}{\left[(2\pi)^4\right]^2} \; P_\mu \, 
\vec{\rho}_\mu(Q)\cdot \vec{\pi}(-P-\frac{Q}{2})\times\vec{\pi}(P-\frac{Q}{2}).
\eeq
which is equivalent to the more standard form
\beq
S\left[\rho\pi\pi\right]=-g_{\rho\pi\pi}\int d^4x \; 
\vec{\rho}_\mu(x)\cdot \vec{\pi}(x)\times\partial_\mu\vec{\pi}(x).
\label{pointaction}
\eeq
With the parameterized BS amplitude  
\mbox{$\Gamma_V(q;P) \propto exp(-q^2/a^2)$}, we take $a=0.194~{\rm GeV}^2$
to reproduce the value $g_{\rho\pi\pi}^{\rm expt}=6.05$ as inferred from the 
experimental $\rho\rightarrow\pi\pi$ decay width of $151~{\rm MeV}$.
The calculated form factor 
$F_{\rho\pi\pi}(Q^2)$ is displayed in Fig.~\ref{rho} for timelike and
spacelike momenta in the vicinity of the mass-shell.
A previous study\cite{rpp87} within the GCM  made the approximation
$g_{\rho\pi\pi} \approx F_{\rho\pi\pi}(P=Q=0)$  to avoid the occurrence of
complex momenta in the arguments of propagators and vertex functions in the 
quark loop.   From Fig.~\ref{rho} it is evident this approximation can
underestimate the value by almost a factor of two.

The pion loop mechanism to be considered later requires the $\rho\pi\pi$ 
vertex for off-shell pion momenta.  Only the form factor 
$F_{\rho\pi\pi}$ is relevant due to the transverse $\rho$ condition.  
Results from numerical evaluation of Eq.~(\ref{vrpp})  over the relevant 
range of $P^2$ and $Q^2$ for \mbox{$P\cdot Q =0$} are displayed 
in Fig.~\ref{rho_curves}.  
For later use we parameterize the results in the form 
\beq
F_{\rho\pi\pi}(P^2,Q^2)
\approx f_{\rho\pi\pi}(Q^2) e^{-P^2/\lambda_{\rho}^2(Q^2)} .
\label{rhov}
\eeq
The constraint  \mbox{$P\cdot Q=0$} is applied to facilitate subsequent use 
in the $\pi$ loop integral and it is exact at the important on-mass-shell 
point for both pions.   
The functions $f_{\rho\pi\pi}(Q^2)$ and $\lambda_\rho(Q^2)$ are chosen to 
provide a fit to
the calculated $\rho$ momentum dependence of the vertex to within $1\%$
for $Q^2$ between $-m_\rho^2$ and $0.6~{\rm GeV}^2$.  
We use
\mbox{$f_{\rho\pi\pi}(Q^2)=$} \mbox{$\sum_{n=0}^{2} a_n z^n$} 
and \mbox{$\lambda_\rho(Q^2)=$} \mbox{$\mu \, \sum_{n=0}^{6} b_n z^n$} 
with \mbox{$z=Q^2/\mu^2$} and \mbox{$\mu =1~{\rm GeV}$}.  The parameters are  
\mbox{$(a_0,a_1,a_2)=$} \mbox{$(3.04,-5.20,2.37)$} 
and \mbox{$(b_0,\ldots,b_6)=$} 
\mbox{$(1.28,0.252,0.110,0.171,1.08,1.80,0.865)$}.

The $\omega\pi\pi$ interaction term from the effective action 
of Eq.~(\ref{mesonaction}) is  
\beq
S[\omega\pi\pi]=-Tr\left[S_0 i\gamma_\mu \omega_\mu \Gamma_V
(S_0 i\gamma_5 \vec{\tau}\cdot\vec{\pi}\Gamma_{\pi})^2 \right] .
\eeq
This is identical in form to the 
$\rho\pi\pi$ interaction except for the isospin structure.   
If the only isovector quantities involved here are the pion fields, the 
resulting expression is just Eq.~(\ref{srpp}) but with isospin structure 
\mbox{$\vec{\pi} \cdot \vec{\pi}$}. The symmetry properties of 
$\Lambda_\mu^\omega(P,Q)$ then show that the result is identically zero as 
required  
by G-parity conservation.   With  extra isovector components
provided by the quark propagators,  the G-parity violating result is 
\beq
\hat{S}_3[\omega\pi\pi] = i \int \frac{d^{4}P,Q}{[(2\pi)^4]^2} \;
 \omega_{\nu}(Q) \; \hat{3} \cdot \vec{\pi}(-P-Q/2) \times  \vec{\pi}(P-Q/2)
\; \Lambda^{\omega}_{\nu}(P;Q)  ,
\label{opp}
\eeq
where the vertex is
\beq
\Lambda^\omega_\mu(P,Q)=\int \frac{d^4k}{(2\pi)^4}
\Gamma_V(k+\frac{P}{2};Q) \Gamma_\pi(k+\frac{Q}{4};-P-\frac{Q}{2})
\Gamma_\pi(k-\frac{Q}{4};P-\frac{Q}{2}) N_\mu(k,P,Q)  .
\eeq
Here, to lowest order in isospin breaking,  the loop trace is proportional to 
$\delta m $ and is given by  $N_\mu = N_\mu^{100} -N_\mu^{001} +N_\mu^{010}$, 
where
\beq
N_\mu^{abc}(k,P,Q)={\rm tr}\left[
S_0^a(k+\frac{P}{2}+\frac{Q}{2}) \, i\gamma_\mu  \,
S_0^b(k+\frac{P}{2}-\frac{Q}{2}) \, i\gamma_5  \,
S_0^c(k-\frac{P}{2}) \, i\gamma_5  \right] ,
\eeq
with the isospin components of the propagators defined by 
\mbox{$S_0 = S_0^0 + \tau_3 S_0^1$}. 

The symmetry properties of $\Lambda_\mu^{\omega}(P,Q)$ are identical 
to those for $\Lambda_\mu^{\rho}(P,Q)$  and hence the general form is
\beq
\Lambda_\mu^{\omega}(P,Q)=-P_\mu \; F_{\omega\pi\pi}(P^2,Q^2,(P\cdot Q)^2)
-Q_\mu P\cdot Q \; H_{\omega\pi\pi}(P^2,Q^2,(P\cdot Q)^2).
\eeq
At the triple on-mass-shell point, numerical evaluation of $F_{\omega\pi\pi}$ 
yields the coupling constant $g_{\omega_I\pi\pi}=0.105$.  This is almost
five times larger than the value $0.0236$ recently obtained in Ref.~\cite{SS96}.
The $2\pi$ width of the physical $\omega$
is $0.185~{\rm MeV}$ which implies $g_{\omega\pi\pi}^{expt}=0.21$.  The 
theoretical quantity $g_{\omega_I\pi\pi}$ obtained so far refers 
to the pure isospin $\omega_I$ state and a small admixture of $\rho_I$ can 
make a significant contribution to the physical $\omega \rightarrow 2\pi$ 
decay.  We will return to this topic later.  
The calculated off-shell form factor $F_{\omega\pi\pi}(P^2,Q^2)$ is displayed 
in Fig.~\ref{omega_curves} for \mbox{$P\cdot Q =0$}.
A common assumption is that the $\rho\pi\pi$ and $\omega\pi\pi$ 
form factors simply scale by the ratio of their coupling constants   
so that the pion loop $\rho^0-\omega$ self-energy may be estimated
by scaling the corresponding $\rho$ self-energy~\cite{R,OPTWrev}.  From 
Figs.~\ref{rho_curves} and~\ref{omega_curves} the present calculations show
that this can be qualitatively reliable for momenta where the form factors are
significant. The departures that do occur are due to the different momentum 
dependence of the isoscalar and isovector components of the quark propagator. 
 
For the subsequent pion loop calculation we parameterize the $\omega\pi\pi$
form factor in the same way as in Eq.~(\ref{rhov}).  That is, 
\mbox{$f_{\omega\pi\pi}(Q^2)=$} \mbox{$\sum_{n=0}^{2} a_n z^n$}  and 
\mbox{$\lambda_\omega(Q^2)=$} \mbox{$\mu \, \sum_{n=0}^{6} b_n z^n$}  with 
\mbox{$z=Q^2/\mu^2$} and \mbox{$\mu =1~{\rm GeV}$}.   The parameters are
\mbox{$(a_0,a_1,a_2)=$} \mbox{$(7.07,-6.91,3.21)\times 10^{-2}$} and 
\mbox{$(b_0,\ldots,b_6)=$} 
\mbox{$(118,6.76,-1.72,-0.391,-0.172,0.109,0.140)\times 10^{-2}$}.

\section{The $\rho-\omega$ Sector up to One-Pion-Loop}
\label{sec6}

The  relevant low-order terms in the effective action for $\pi,\rho,\omega$ 
that we have produced by integrating out the quarks are  
\beq
\hat{S}\left[\pi,\rho,\omega\right] = \hat{S}_2\left[\rho,\omega\right]
+\hat{S}_2\left[\pi\right] + \hat{S}[\rho\pi\pi] +\hat{S}[\omega\pi\pi] 
+ \cdots  .
\label{sql}
\eeq
Here the second-order terms $\hat{S}_2\left[\rho,\omega\right]$  and
$\hat{S}_2[\pi]$ represent respectively the $\rho-\omega$ sector containing the 
quark-loop mixing mechanism and the free pion sector. 
To integrate out the pion fields, it is convenient to first combine the
last three terms of Eq.~(\ref{sql}) so that it becomes
\beq
\hat{S}\left[\pi,\rho,\omega\right]=\hat{S}\left[\rho,\omega\right]
+\frac{1}{2}\int 
\frac{d^4P,P'}{(2\pi)^8} \, \pi_i(P')D^{-1}_{ij}(P',P)\pi_j(P) .
\eeq
The three terms in
\beq
D_{ij}^{-1}(P',P)=\Delta^{-1}_{ij}(P',P)+V_{ij}(P',P)+W_{ij}(P',P)
\eeq
correspond to the free pion term given by
\beq
\Delta^{-1}_{ij}(P',P)=(2\pi)^4\delta_{ij}\delta^4(P'+P)\Delta^{-1}_\pi(P^2) ,
\eeq
the $\rho\pi\pi$ interaction term given by
\beq
V_{ij}(P',P)=2i\epsilon_{ijk} \, \rho_\mu^k(-P'-P)  \;
\Lambda_\mu^\rho(-\frac{P'+P}{2};-P'-P) ,
\eeq
and the $\omega\pi\pi$ interaction term given by
\beq
W_{ij}(P',P)=2i\epsilon_{ij3} \, \omega_\mu(-P'-P)  \;
\Lambda_\mu^\omega(-\frac{P'+P}{2};-P'-P).
\eeq
Integration over the pion fields allows a new effective action 
$\hat{{\cal S}}[\rho,\omega]$ to be identified from
\beq
Z=N \int D\rho D\omega  D\pi \, exp\left( -\hat{S}[\pi,\rho,\omega] \right)
= N' \int D\rho D\omega \, exp\left( -\hat{{\cal S}}[\rho,\omega] \right) 
\eeq
by using the functional integral result
\beq
\int D\pi \, exp\left( -\frac{1}{2}\int \frac{d^4P,P'}{(2\pi)^8} \,
\pi_i(P')D^{-1}_{ij}(P',P)\pi_j(P) \right)
=exp\left( -\frac{1}{2} {\rm TrLn} \; D^{-1}\right)  .
\eeq
We absorb the field-independent term \mbox{$exp(-\frac{1}{2} {\rm TrLn} \; 
\Delta^{-1} )$} into the normalization constant to arrive at
\beq
\hat{{\cal S}}\left[\rho,\omega\right]=\hat{S}_2\left[\rho,\omega\right]
+\frac{1}{2} {\rm TrLn} (1+\Delta(V+W)).
\label{spiloop}
\eeq
The second term here gives the vector meson coupling to all orders
produced by a single pion loop.  

The quadratic part of the second term of Eq.~(\ref{spiloop}) adds
a contribution to the vector meson inverse propagator to that given previously
by Eq.~(\ref{qlinvpr}) at the quark loop level.  In the transverse 
$\rho^0-\omega$ sector, the net result for the inverse propagator is
\beq
{\cal D}^{-1}(P^2) = \left( \begin{array}{cc} 
P^2+m_V^2 +\Pi^{\rho\rho}(P^2)    & \Pi^{\rho\omega}(P^2) \\
\Pi^{\rho\omega}(P^2) & P^2+m_V^2 +\Pi^{\omega\omega}(P^2) \end{array} \right) ,
\label{totinvpr}
\eeq
where the net mixing amplitude is \mbox{$\Pi^{\rho\omega}(P^2)= 
\Pi^q(P^2)+\Pi^\pi(P^2)$}.
The additional terms produced by the pion loop are the diagonal elements
$\Pi^{\rho\rho}(P^2)$ and  $\Pi^{\omega\omega}(P^2)$, and the mixing term 
$\Pi^\pi (P^2)$.   From  Eq.~(\ref{spiloop}),
the self-energy contribution $\Pi^{\rho\rho}(P^2)$ is given by the transverse  
component of
\beq
\Pi^{\rho \rho}_{\mu \nu}(P)= -4 \int \frac{d^4 q}{(2 \pi)^4} \Delta_\pi(q_+)
\Delta_\pi(q_-) \Lambda^{\rho}_{\mu}(q;-P) \Lambda^{\rho}_{\nu}(q;P)  ,
\label{rhose}
\eeq
where \mbox{$q_\pm = q \pm P/2$}, and the summation over pion isospin labels
has been carried out.   This contribution is the same for each isospin 
component of the $\rho$.  Similarly, the mixed $\rho^0-\omega$ self-energy 
contribution generated by the pion loop is given by  
\beq
\Pi^{\pi}_{\mu \nu}(P)= -4 \int \frac{d^4 q}{(2 \pi)^4} \Delta_\pi(q_+)
\Delta_\pi(q_-) \Lambda^{\rho}_{\mu}(q;-P) \Lambda^{\omega}_{\nu}(q;P),
\label{mixse}
\eeq
and $\Pi^\pi(P^2)$ is the transverse component of that. 
The $\omega$ self-energy contribution is
obtained from Eq.~(\ref{rhose}) by the replacement  of $\Lambda^\rho$ by
$\Lambda^\omega$.  It is quadratic in the small symmetry breaking mechanism 
and is ignored here.

In the absence of mixing,  the real part of 
$~\Pi^{\rho\rho }(P^2)$ generates  a mass shift for the isospin 
eigenstate $\rho_I$.  Since the corresponding quantity in the $\omega_I$ 
channel is much smaller, and neglected in this work, this mass shift 
represents the $\rho-\omega$ mass splitting.   
For timelike momenta such that $P^2 \leq -4 m_\pi^2$, both
$~\Pi^{\rho \rho }(P^2)$ and $~\Pi^{\pi }(P^2)$ have imaginary parts associated 
with the decay $\rho \rightarrow 2 \pi$.   The imaginary  
part of $~\Pi^{\rho\rho }(P^2)$ produces the $2\pi$ width of the $\rho_I$ 
state.

These observations are, in principle, modified by the presence of nonzero
off-diagonal elements in Eq.~(\ref{totinvpr}).  To see this, we construct
the eigenvalues of the inverse meson propagator, ${\cal D}^{-1}(P^2)$. These 
are, to lowest order in the 
total mixing amplitude, given by the expressions
\mbox{$\lambda_\rho(P^2) = P^2+m_V^2 +\Pi^{\rho\rho}(P^2)(1+\epsilon^2(P^2))$},
and 
\mbox{$\lambda_\omega(P^2) = P^2+m_V^2 -\Pi^{\rho\rho}(P^2)\epsilon^2(P^2))$},
where 
\beq
\epsilon(P^2) = - \frac{\Pi^{\rho\omega}(P^2)}{\Pi^{\rho\rho} (P^2)}  
\label{epsilp2}
\eeq
is a natural measure of off-diagonal coupling.  
Since the typical size is $|\epsilon| \approx 0.03$ in the mass-shell region, 
we shall ignore the second order mixing effects on the mass shifts. 
In the diagonal basis we therefore have, through first order in mixing,
\beq
{\cal D}^{-1}(P^2) = \left( \begin{array}{cc} 
P^2+m_\omega^2 +\Pi^{\rho\rho}(P^2)    &   0   \\
       0           & P^2+m_\omega^2  \end{array} \right) ,
\label{diainvpr}
\eeq
where we have made the identification $m_\omega = m_V$.  
We identify the mass of the (mixed) 
$\rho$ eigenstate from the position of the zero in the real part of the
upper eigenvalue in Eq.~(\ref{diainvpr}).  That is, 
\beq
m_\rho^2 =m_\omega^2+ {\rm Re} \, \Pi^{\rho\rho}(-m_\rho^2) .
\label{rhomass}
\eeq
With the definition 
\mbox{$\hat{\Gamma}(P^2) = -{\rm Im}\, \Pi^{\rho\rho}(P^2)/m_\rho$}, 
the inverse propagator in the $\rho$ channel can be written as
\beq
{\cal D}^{-1}_\rho(P^2) = (P^2 + m_\rho^2 -i m_\rho \Gamma(P^2)) \, Z_2(P^2)  ,
\label{rinvpr}   
\eeq
where \mbox{$\Gamma(P^2)=\hat{\Gamma}(P^2)/Z_2(P^2)$}.  The function 
$Z_2(P^2)$ arises from the momentum dependence of the real part of the pion 
loop self-energy.  The on-mass-shell value, which is to be absorbed into the 
$\rho$ field as a renormalization constant, is given by
\beq
Z_2(-m_\rho^2) = 1 + {\rm Re}\, \Pi^{\rho\rho'}(-m_\rho^2)  ,
\label{z2}
\eeq
with the prime superscript denoting differentiation with respect to the 
argument.  We choose to absorb the function $\sqrt{Z_2(P^2)}$ into the 
field $\rho_\mu(P)$ so that the resulting propagator has the conventional
Breit-Wigner form.  The  physical width is given by
\beq
\Gamma_\rho = -\frac{ {\rm Im}\, \Pi^{\rho\rho}(-m_\rho^2)}
{m_\rho \, Z_2(-m_\rho^2)}  .
\label{rwid}
\eeq

Our numerical results for the $\rho$ mass shift and width are obtained from 
Eq.~(\ref{rhose}) in the specific form
\beq
\Pi^{\rho\rho}(P^2)=-\frac{4}{3}f_{\rho\pi\pi}^2(P^2) \int 
\frac{d^4q}{(2\pi)^4}
\frac{( q^2- (q \cdot P)^2/P^2 ) e^{-2 q^2/\lambda_{\rho}^2(P^2)}}
{\left[(q+\frac{P}{2})^2+m_\pi^2-i\epsilon\right]
\left[(q-\frac{P}{2})^2+m_\pi^2-i\epsilon\right]} ,
\eeq
where the parameterized $\rho\pi\pi$ vertex functions from Eq.~(\ref{rhov})
have been used.  After reduction to a two-dimensional integral it is convenient
to employ the Feynman combination of denominators to obtain
\beq
\Pi^{\rho\rho}(P^2)=-\frac{4}{3}f_{\rho\pi\pi}^2(P^2) 
\int_{-\frac{1}{2}}^{+\frac{1}{2}} d\alpha
\int_0^\infty ds s^2
\int_{-1}^{+1} dz \sqrt{1-z^2}
\frac{e^{-2(s-2\alpha z s P+\alpha^2 P^2)/\lambda^2_\rho(P^2)}}
{\left[s+m_\pi^2+\frac{P^2}{4}-\alpha^2 P^2-i\epsilon\right]^2}.
\eeq
The numerical results for ${\rm Re}\, \Pi^{\rho\rho}(P^2)$ and ${\rm Im}\, 
\Pi^{\rho\rho}(P^2)$ are shown in Fig.~\ref{rho_pl}.  
The self-consistent solution for $m_\rho$ produced by Eq.~(\ref{rhomass}) 
is displayed in Fig.~\ref{mcorr_comp} in the following way.
The quantity $-(P^2+m_\omega^2)$ is plotted as a long dash line and its 
intercept with the solid line representing ${\rm Re}\, \Pi^{\rho\rho}(P^2)$ 
identifies the mass shell point.  We use $m_\omega=m_V=782~{\rm MeV}$  and
find \mbox{$m_\rho=761~{\rm MeV}$} giving
\mbox{$m_\rho - m_\omega = -21~{\rm MeV}$}.   
The experimental value is $-12.0\pm 0.8$ MeV.  
At this mass-shell point we find 
\mbox{$\Pi^{\rho\rho}(-m_\rho^2) = -0.031 - 0.11 i~{\rm GeV}^2$}. 
The calculated field renormalization constant from Eq.~(\ref{z2}) is 
\mbox{$Z_2(-m_{\rho}^2) = 0.90$}.  Then Eq.~(\ref{rwid}) yields the width
\mbox{$\Gamma_\rho = 156~{\rm MeV}$} while the experimental value is 
$151~{\rm MeV}$.   

Fig.~\ref{mcorr_comp} also illustrates the contribution to the mass splitting
result made by the momentum dependence of the meson self-energies  
generated by the $\bar{q}q$ substructure.  This is most marked for the pion
and enters the pion loop integral in Eq.~(\ref{rhose}) through the factors 
$Z_\pi^{-1} (q_\pm^2)$
which are contributed by each pion propagator.  (In our calculational procedure
these factors have been moved into the vertex functions for convenience.)  If 
each 
function $Z_\pi$ is held constant at its mass-shell value 
(as would be the case for a structureless pion), the full calculation 
in Fig.~\ref{mcorr_comp} (solid line) becomes the dot-dashed line.  The 
intercept then indicates that the mass shift would be essentially zero.  In
contrast to this, the short dashed line indicates that the influence of
the corresponding quantity $Z_1(P^2)$, from the substructure of the $\rho$,  
is quite negligible here.  This is because the range of variation in $P^2$ 
is very small.

In a parallel manner, the pion loop contribution to the 
transverse $\rho^0-\omega$ mixing amplitude is calculated from
\beq
\Pi^\pi (P^2)=-\frac{4}{3}f_{\rho\pi\pi}(P^2)f_{\omega\pi\pi}(P^2) 
\int \frac{d^4q}{(2\pi)^4}
\frac{  (q^2-(q\cdot P)^2/P^2) e^{-q^2/\lambda^2(P^2)}  }
{\left[(q+\frac{P}{2})^2+m_\pi^2-i\epsilon\right]
\left[(q-\frac{P}{2})^2+m_\pi^2-i\epsilon\right]}
\eeq
where $\lambda^{-2}=\lambda_\rho^{-2}+\lambda_\omega^{-2}$.
The results are shown in Fig.~\ref{mix_pl}.  An experimental
value~\cite{coon} for the total real mixing amplitude is also displayed. 
The pion loop mechanism produces a significantly smaller real amplitude than 
does the quark-loop mechanism and does not have a node near zero momentum.   
In Fig.~\ref{mix_comp} we illustrate the contributions to ${\rm Re}~\Pi^\pi$ 
made by the dynamical nature of the meson
self-energy functions.  When the pion quantities $Z_\pi (q_\pm^2)$
are held fixed at the mass-shell value, the full calculation (solid line) 
reduces to the dot-dashed line.   As with the $\rho$ self-energy,
the dynamical $\bar{q}q$ substructure of the  pion has a considerable 
influence on the pion loop mixing mechanism near the vector 
meson mass-shell.  The difference
between the solid and dashed lines shows that the corresponding
vector meson substructure plays a lesser role. 
We compare the quark loop (dashed line) and pion loop (dotted line) 
contributions to the real mixing amplitude in Fig.~\ref{sum}.  
The solid line is the sum of the two.    At the mass-shell, the 
quark loop dominates with the pion-loop contributing about
$20\%$ of the total.  
For spacelike momenta, the pion loop contribution is negligible.

The change of basis associated with the diagonal form of the inverse 
propagator given in Eq.~(\ref{diainvpr}) produces physical fields given by
\beq
\left( \begin{array}{c}  \rho \\  \omega  \end{array} \right) 
 = \left( \begin{array}{cc} 
     1    &   -\epsilon   \\
  \epsilon  &   1  \end{array} \right) 
\left( \begin{array}{c}  \rho_I \\  \omega_I  \end{array} \right)  .
\label{diafids}
\eeq
Currents that couple to $\rho_I$ and $\omega_I$ are mixed in the same way.
Thus the resonant $\rho$ and $\omega$ contribution to the time-like pion
charge form factor is given by
\begin{equation}
F_{\pi}^R(P^2) = F_{\omega\pi\pi} (P^2) {\cal D}_\omega (P^2) 
f_{\omega\gamma}(P^2) + F_{\rho\pi\pi} (P^2) {\cal D}_\rho (P^2) 
f_{\rho\gamma}(P^2)  ,
\end{equation}
where \mbox{${\cal D}_\omega (P^2) = [P^2 + m_\omega^2]^{-1}$} and 
\mbox{${\cal D}_\rho (P^2) = [P^2 + m_\rho^2 - im_\rho \Gamma(P^2)]^{-1}$}. 
The $2\pi$ form factors of the physical vector mesons are
\begin{eqnarray}
F_{\omega\pi\pi} (P^2) = F_{\omega_I\pi\pi} (P^2) + \epsilon(P^2)
F_{\rho_I\pi\pi} (P^2)  \nonumber   \\
F_{\rho\pi\pi} (P^2) = \frac{1}{ \sqrt{Z_2(P^2)} } \left(F_{\rho_I\pi\pi} (P^2) 
- \epsilon(P^2) F_{\omega_I\pi\pi} (P^2)\right) . 
\label{physff}
\end{eqnarray}
The form factors in the isospin basis have been described in 
Sec.~\ref{sec5}.  The dependence of the electromagnetic coupling strengths 
$f_{\rho\gamma}$ and $f_{\omega\gamma}$ on the mixing amplitude must be 
accounted for in a fit to data. (See, for example Ref.~\cite{MOW96}.)  Here we
limit our consideration to aspects of the  $2\pi$ decay.  Since the mixing 
amplitude $\epsilon$ is part of a small second order effect in 
$F_{\rho\pi\pi} (P^2)$, it is the resonant $\omega$ contribution, and in
particular the physical coupling constant 
\mbox{$g_{\omega\pi\pi} = F_{\omega\pi\pi}(-m_{\omega}^2)$}, which carries
the leading information on the mixing amplitude.  From Eq.~(\ref{physff})
we may write
\beq
g_{\omega \pi \pi} = g_{\omega_I \pi \pi} + \epsilon(-m_{\omega}^2)
~F_{\rho_I \pi \pi}(-m_{\omega}^2)  ,
\label{go}
\eeq
by making  use of the fact that $\omega$ and $\omega_I$ have 
the same mass at the present level of approximation.  Given that 
$F_{\rho_I \pi \pi}(-m_{\omega}^2)$ is well determined by the known
$g_{\rho\pi\pi}$, knowledge of $g_{\omega_I \pi \pi}$ is necessary before
$\epsilon$ can be deduced from a value of $g_{\omega\pi\pi}$ extracted from 
data analysis.   

For the calculated total mixing amplitude at the $\omega$ mass shell, we find 
\mbox{$\Pi^{\rho\omega} (-m_{\omega}^2) =$} 
\mbox{$-(2.6 +2.3i) \times 10^{-3}~{\rm GeV}^2$}.   The
real part is substantially below the experimental value~\cite{coon} of
$-(4.52\pm 0.6) \times 10^{-3}~{\rm GeV}^2$ from analysis of the 
$e^+e^- \rightarrow \pi^+ \pi^- $ data.  This latter value relies upon  
assumptions~\cite{R} in the analysis that lead to the cancellation of the 
direct $\omega_I \rightarrow \pi \pi$ mechanism with 
${\rm Im}~\Pi^{\rho\omega}$.  A recent  re-analysis~\cite{MOW96}  has
pointed out that the corrections to these assumptions lead to an
increase of about a factor of three in the uncertainty of the extracted 
real mixing amplitude in the absence of theoretical limits on 
$g_{\omega_I \pi\pi}$.   Our result for 
${\rm Re}~\Pi^{\rho\omega} (-m_{\omega}^2)$ is just within such an allowed 
range.  The pion loop contribution to the $\rho$ self-energy is found to be
\mbox{$\Pi^{\rho\rho} (-m_{\omega}^2) =$}
\mbox{$ -(0.31 +1.1i) \times 10^{-1}~{\rm GeV}^2$}
giving a mixing parameter \mbox{$|\epsilon(-m_{\omega}^2)| = 0.0303$}, 
\mbox{${\rm arg}(\epsilon) = 147.2^\circ$} from Eq.~(\ref{epsilp2}).  

The 
influence of the direct coupling term $g_{\omega_I \pi \pi}$ within the 
present model can be assessed as follows.  With Eq.~(\ref{epsilp2}) 
for $\epsilon$,  Eq.~(\ref{go}) becomes
\beq
g_{\omega \pi \pi} = - \frac{ {\rm Re} ~\Pi^{\rho\omega}}{\Pi^{\rho\rho}} 
~g_{\rho_I \pi \pi} 
+ \frac{ g_{\omega_I\pi\pi} \Pi^{\rho\rho} - i {\rm Im} ~\Pi^{\rho\omega} 
g_{\rho_I\pi\pi} }
       {\Pi^{\rho\rho}}   ,
\label{go2}
\eeq
where we write 
\mbox{$F_{\rho_I \pi\pi}(-m_\omega^2) \approx g_{\rho_I \pi\pi}$} since
the shifted mass-shell point is a minor correction to this quantity.  
Although the amplitudes $\Pi^{\rho\omega}(P^2)$ and $\Pi^{\rho\rho}(P^2)$ 
here are to be evaluated at $-m_{\omega}^2$, the arguments below are not 
altered significantly by changes in the mass-shell point compatible with the 
$\rho-\omega$ mass difference.
The second term above is usually eliminated by making two 
assumptions~\cite{R}. Firstly, ${\rm Re} \Pi^{\rho\rho}$ is taken to be zero. 
Apart from small adjustments due to 
the momentum dependence, this ignores the $\rho-\omega$ mass
difference given in Eq.~(\ref{rhomass}). Secondly, ${\rm Im} ~\Pi^{\rho\omega}$ 
is assumed to be ${\rm Im} ~\Pi^{\rho\rho}$ scaled down by 
$g_{\omega_I\pi\pi}/g_{\rho_I\pi\pi}$. 
This would be exact in the present model if the $\rho\pi\pi$ and 
$\omega\pi\pi$ form factors were characterized by the same momentum dependent 
shape.  As we have seen from Figs.~\ref{rho_curves}  and \ref{omega_curves}, 
this is only true in a qualitative sense.  As a result of the relevant loop 
integrals for the self-energies, there is an  
approximate scaling behavior evident in Figs.~\ref{rho_pl} and \ref{mix_pl}.
Quantitatively, we find this second assumption to be in error by
about $17\%$ at the mass shell.  To assess the net effect of 
these approximations, Eq.~(\ref{go2}) may be expressed as
\beq
g_{\omega \pi \pi} = - \frac{ ({\rm Re}~\Pi^{\rho\omega}) \; {\cal C} } 
{i{\rm Im}~\Pi^{\rho\rho}} ~g_{\rho_I \pi \pi} 
\label{go3}
\eeq
where the denominator is determined by the $\rho$ mass and width via 
Eq.~(\ref{rwid}).   The departure of the correction factor ${\cal C}$ from its 
conventional value of unity is a measure of the accuracy of the
approximations mentioned above.  Within the present model we find 
\mbox{$| {\cal C}| = 0.78$} and \mbox{${\rm arg}( {\cal C} ) = 25^\circ$}.
A significant part of this correction is generated by  
$g_{\omega_I\pi\pi} {\rm Re} ~\Pi^{\rho\rho}$ and thus is a measure of the 
competing influence of the usually neglected direct 
$\omega_I \rightarrow \pi \pi$ process.  The trend is for a deduced value 
of ${\rm Re}~\Pi^{\rho\omega}$ to be some $20\%$ larger than what would 
otherwise be needed for the same $g_{\omega\pi\pi}$.

\section{Discussion}
\label{sec7}

In this work, we have found that a number of properties of the $\rho-\omega$
system can be described by a phenomenological treatment of dressed quark
degrees of freedom as implemented by an effective field theory model  
applied up to one pion loop.   A distinguishing feature of the present 
approach is that the finite size $\bar{q}q$ substructure of the mesons is
included.   This substructure for the pion is correlated with the 
momentum-dependent scalar self-energy dressing of the quark through chiral
symmetry considerations.   The resulting dynamical nature of the mass 
function in the pion propagator is included in the calculated pion loop
contributions to both the mixed $\rho^0-\omega$ self-energy  and the 
diagonal self-energy in the $\rho$ channel.  The latter produces a reasonable
value for the $\rho-\omega$ mass difference if the influence of the 
$\bar{q}q$ substructure effects on the pion propagator are included.  If we
use the point particle limit of the pion propagator, this mass difference
becomes negligible.
The total $\rho^0-\omega$ mixing here consists of a quark loop and a pion
loop part and the latter is found to contribute $20\%$ at the mass-shell.
For spacelike momenta appropriate to mixed $\rho-\omega$ exchange for the NN
force, the pion loop mixing contribution is not significant and the total 
mixing amplitude remains much too small to explain the empirical CSB NN force.

The calculated values of $g_{\rho\pi\pi}$ and $m_\omega$ are used to fix
two parameters that we have introduced to minimally add the vector mesons to
a previously developed parameterization of pion physics that is the basis of 
the quark-level approach we employ.  
The isospin breaking quantities calculated here, i.e.
the $\rho^0-\omega$ mixing amplitude and the coupling constant 
$g_{\omega_I \pi\pi}$, scale in an approximately linear way with the quark
bare mass difference and we have presented our results for the typical value
\mbox{$\delta m = -4~{\rm MeV}$}. 

We do not consider the contribution of $\rho^0-\omega$ mixing to the meson
masses because it is an effect of second order in the mixing amplitude.  For 
a similar reason we do not consider the one pion loop correction to 
$m_{\omega}$.  In our pion loop calculations, we 
have used the empirical value of $g_{\rho\pi\pi}$ to set the strength of the
$\rho^0_I\pi\pi$ vertex even though the empirical coupling constant refers
to the physical $\rho^0$ which is a linear combination of $\rho^0_I$ and 
$\omega_I$.  For a calculation that is strictly of lowest
order in the mixing amplitude, one should use the model results for
the mixing to deduce a value for $g_{\rho_I \pi\pi}$  and iterate the
calculation until there is self-consistency.  However consideration of 
Eq.~(\ref{physff}) shows that the correction is a small second order effect
and we do not make it.  

In this work the coupling constant $g_{\omega_I \pi\pi}$ is found to be
\mbox{$g_{\omega_I \pi\pi}= 0.105$}.  We examine the correction that this 
direct process brings to the relation between the 
\mbox{$\omega\rightarrow\pi\pi$} decay and $\rho^0-\omega$ mixing amplitude.
The trend is for the necessary real mixing amplitude to be some $20\%$ larger
than what would otherwise be the case.  There is also a phase consequence and
specific details can be found in Eq.~(\ref{go3}).

In a future work some of the approximations we make could be removed and 
some reduction could be sought in the degree of phenomenology employed. 
Although the strength of the single scalar function we use for the vector
meson BS amplitude is determined by the mass-shell normalization internal
to the model, the range of the gaussian form is described by a free parameter 
constrained by the calculated $g_{\rho\pi\pi}$.  We believe the result to be  
physically sensible because, in this same general framework, it leads to
the prediction~\cite{M95,T96}  \mbox{$g_{\gamma\pi\rho} = 0.50$} which is to 
be compared to \mbox{$g_{\gamma\pi\rho}^{expt} = 0.504$}.  Nevertheless, 
it is of interest to check these results with a vector meson amplitude that 
solves a BS equation and allows more than just the canonical Dirac matrix 
covariant.   Such numerical solutions are available~\cite{JM93}.
However it is difficult to accommodate the consequences of the complex values 
of quark loop momenta produced by mass-shell values of meson momenta in the
Euclidean metric of this model.  An ansatz for treatment of the ladder BS 
equation developed in recent works~\cite{sep96,CG95} goes a long way to 
overcoming this obstacle.  Those works provide conveniently expressed BS 
amplitudes with parameters set by pion and kaon properties.   One such 
approach~\cite{sep96} allows a number of independent covariants selected 
by the dynamics and the predicted amplitudes for $\rho$ and $\omega$ do 
confirm  that the single canonical covariant used in the present work is 
strongly dominant.  The other approach~\cite{CG95} also provides a  
convenient form of the bare mass dependence of parameterized quark 
propagators that is somewhat more general than what we consider here and it 
would be of interest to explore the consequences.

%
\acknowledgements 
This work was supported in part by the National Science Foundation under
Grant Nos. PHY91-13117 and PHY94-14291.  We wish to thank K. Maltman for 
several very helpful discussions. 

\appendix
\section*{Mesons from Quarks}
\label{app}

Hadronization of the quark action of Eq.~(\ref{sgcm})
can be implemented through a functional change of field 
variables in the path integral for the generating functional. The  
meson sector follows directly from a standard 
Fierz reordering of the current-current term of the action in Eq.~(\ref{sgcm}) 
to obtain
\begin{equation}
\frac{1}{2}j_{\mu }^a(x)D(x-y)j_{\mu }^a(y)=-\frac{1}{2}J^{\theta }(x,y)
D(x-y)J^{\theta}(y,x),
\label{jj}
\end{equation}
in terms of bilocal currents 
$J^{\theta }(x,y)= \bar{q}(x)\Lambda ^{\theta }q(y)$.  The $\Lambda ^{\theta }$ 
are direct products  of Dirac spin, $SU(2)$ flavor,   
and $SU(3)$ color matrices, viz.
\begin{equation}
\Lambda ^{\theta }=\frac{1}{2}\left( {\bf 1}_{D},i\gamma
_{5},\frac{i}{\sqrt{2}}\gamma _{\nu },\frac{i}{\sqrt{2}}\gamma _{\nu
}\gamma _{5}\right) \otimes \left( \frac{{\bf 1}_F }{\sqrt{2}},
\frac{ \tau^a }{\sqrt{2}} \right) \otimes \left( \frac{4}{3}{\bf
1}_{C},\frac{i}{\sqrt{3}}\lambda^a\right) .\label{lam}
\end{equation}
We are here interested in the color singlet $\bar{q}q$ sector.  There is
however, a color octet $\bar{q}q$
sector in Eq.~(\ref{lam}).  An alternate color Fierz reordering may be used to
eliminate the color octet $\bar{q}q$ sector in favor of color
anti-triplet $qq$ and $\bar{q}\bar{q}$ pairings which find a natural role
as constituents of baryons~\cite{rtc}.  However only
the color singlet meson modes from Eq.~(\ref{lam}) are of concern here. 
The part of the integrand of the partition function
$Z= N \int D\bar{q}Dq\; exp(-S[\bar{q},q])$ that arises from the right side of 
Eq.~(\ref{jj}) may be expressed as the integral
\begin{equation}
exp \left(\frac{1}{2}(J,DJ)\right) = N' \int D{\cal B} \; 
exp \left[ -\frac{1}{2} \left( {\cal B}, D^{-1}{\cal B} \right)  
- \left( J, {\cal B} \right) \right]
\label{gint}
\end{equation}
where the bracket notation denotes the relevant space-time 
integration and summation over internal degrees of freedom.  The net result
is the reformulation 
\mbox{$Z=$}\mbox{$ N'' \int D{\cal B}D\bar{q}Dq $}
\mbox{$ exp ( -S[\bar{q},q,{\cal B}] ) $} where
the new action has no four fermion term but instead contains
\mbox{$\left( J,{\cal B} \right) $}\mbox{$\equiv \int d^4x,y\; \bar{q}(x) 
\Lambda^{\theta}{\cal B}^{\theta}(x,y) q(y)$}, a linear coupling of 
an appropriate set of bilocal auxiliary fields ${\cal B}^\theta$ to
quarks.  To preserve the symmetries of the original action, it is necessary
that the fields ${\cal B}^{\theta }(x,y)$ have the same
transformation properties as the currents 
$\bar{q}(y)\Lambda ^{\theta }q(x)$.  These auxiliary fields are then 
bosons that take on the dynamics of $\bar{q}q$ correlations. 
For example, the $\omega$ meson of the model will be dominated
by the component associated with \mbox{$\Lambda^\theta \propto i \gamma_\mu$}. 
With the action now quadratic in
the quark fields, they may be integrated out by standard methods.
The partition function then has the representation
$Z={\cal N}\int D{\cal B}\; e^{-S[{\cal B}]}$ in terms of a bosonized action
given by
\begin{equation}
S[{\cal B}]=-\mbox{TrLn}
\left[ \not \! \partial +m +\Lambda ^{\theta }{\cal B}^{\theta }  \right] 
+\int d^4xd^4y\;\frac{ {\cal B}^{\theta}(x,y){\cal B}^{\theta}(y,x) }{2D(x-y)}.
\label{sb}
\end{equation}

An effective action for the propagating meson modes is defined by expansion
in fluctuations about the saddle point of the action
(which is equivalent to the classical vacuum).  From
\mbox{$\delta S[{\cal B}_0] /\delta {\cal B}_0 =0$}, these classical 
configurations given by \mbox{$ {\cal B}^{\theta }_0(r)=$}
\mbox{$D(r){\rm tr}[ \Lambda^{\theta} (\not \! \partial +m +\Lambda^{\theta}
{\cal B}^{\theta}(r) )^{-1} ] $}
are related to nonlocal vacuum condensates and provide a quark self-energy 
\mbox{$\Sigma (r) = \Lambda ^{\theta }{\cal B}^{\theta }_0(r)$}. 
This is equivalent to the ladder Dyson-Schwinger equation in the bare vertex
or rainbow approximation
\begin{equation}
\Sigma (p) = \int \frac{d^{4}q}{(2\pi )^{4}}D(p-q)\frac{\lambda
^{a}}{2}\gamma _{\mu }\frac{1}{i\gamma \cdot q + m + \Sigma
(q)}\frac{\lambda ^{a}}{2}\gamma _{\mu} .\label{sde}
\end{equation}
>From the general form \mbox{$\Sigma(p)= 
i\not \! p\left[ A(p^2)-1\right] +B(p^2)$}, 
the dressed quarks now have a dynamically-generated mass function
\mbox{$M(p^2)=$} \mbox{$(B(p^2)+m)/A(p^2)$}
and become the constituents of the meson modes. 

The meson modes are defined by expansion of the action about the saddle point
configuration of each auxiliary field. That is, one sets
\beq
\Lambda^{\theta}{\cal B}^{\theta}(x,y) = \Sigma(x-y)+
i\gamma_5 \vec{\tau} \cdot \vec{\pi}(x,y) + i\gamma_\mu \omega_\mu(x,y) + 
\cdots  ,
\label{fluct}
\eeq
where $\vec{\pi}(x,y)+\cdots$ are the new field variables, to produce
the meson action from
\begin{equation}
\hat{S}[\pi,\rho,\omega \cdots] = S[ {\cal B}] - S[ {\cal B}_0] .
\label{shat}
\end{equation}
This will consist of terms at the quadratic and higher order level in the 
meson fields.  Contact with effective local
field variables may be made in the following way~\cite{cahill89}.  
For each fluctuation field $\phi$, the free inverse propagator is given by
\mbox{${\cal D}^{-1} = \delta^2 \hat{S}/\delta \phi^2$} at $\phi = 0$.
The free equation of motion ${\cal D}^{-1} \hat\Gamma = 0 $ produces
the ladder Bethe-Salpeter equation for the amplitude $\hat\Gamma$.
The set of eigenfunctions produced by
\mbox{${\cal D}^{-1} \hat{\Gamma}_n = \alpha_n \hat{\Gamma}_n$} may be used 
to expand the bilocal fluctuation fields thereby discretizing the continuous 
internal degree of freedom in terms of known amplitudes, leaving a local 
field as the remaining variable.  Thus, for example, one arrives at the
representation  \mbox{$\pi(q;P) = \sum_n \hat{\Gamma}_n(q;P) \pi_n(P)$}  
where $q$ is the internal momentum conjugate to $x-y$, and $P$, 
conjugate to $(x+y)/2$, is the total momentum of the meson mode described by 
the local variable $\pi_n$.  With truncation to the lowest mass mode
in each case, the result is Eq.~(\ref{mesonaction}).  

\newpage

\begin{figure}
\centerline{\psfig{figure=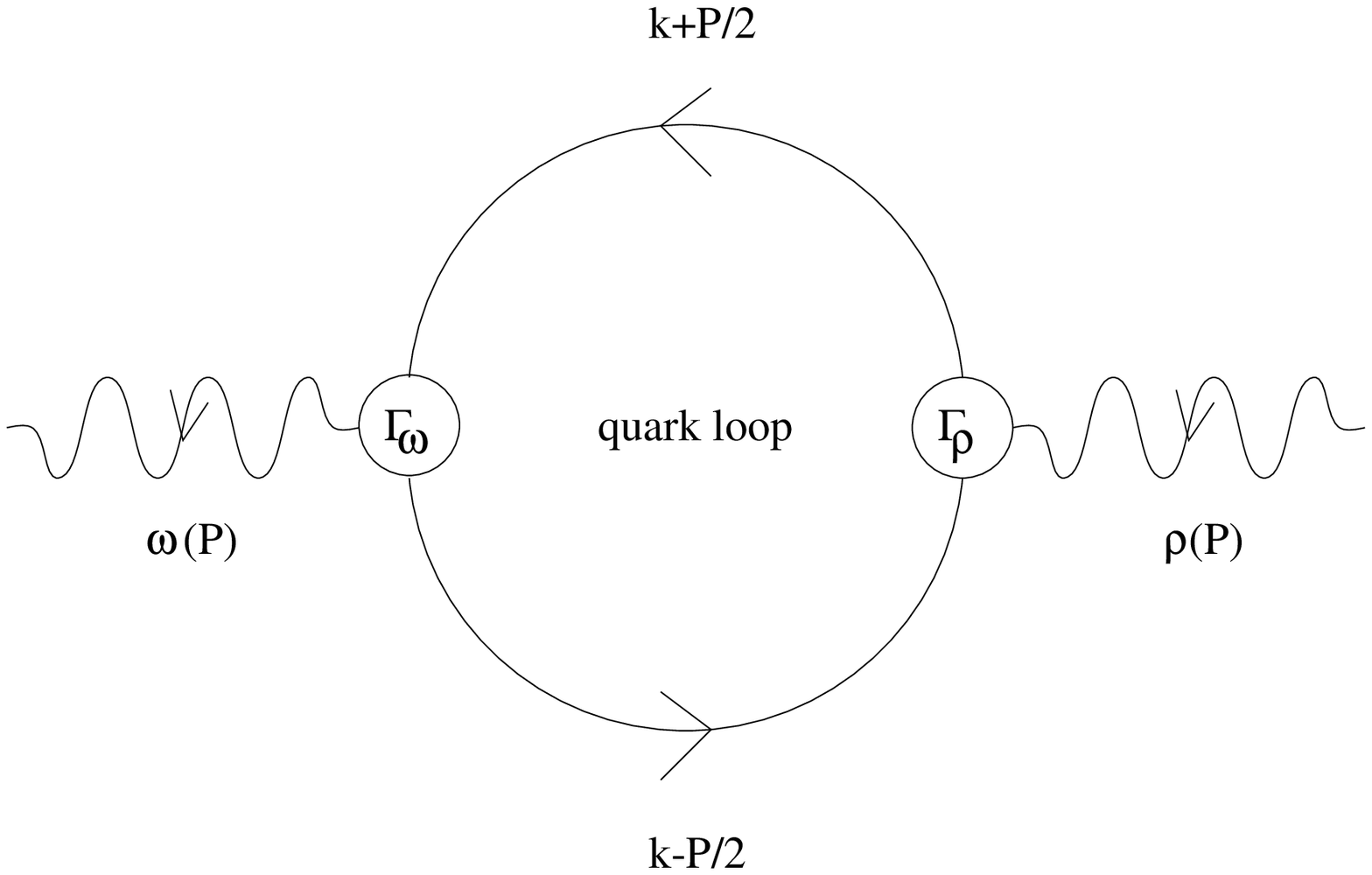,height=7.2in,width=7.2in}}
\caption{The quark-loop mechanism for $\rho\omega$ mixing.\label{qloop}}
\end{figure}

\begin{figure}
\centerline{\psfig{figure=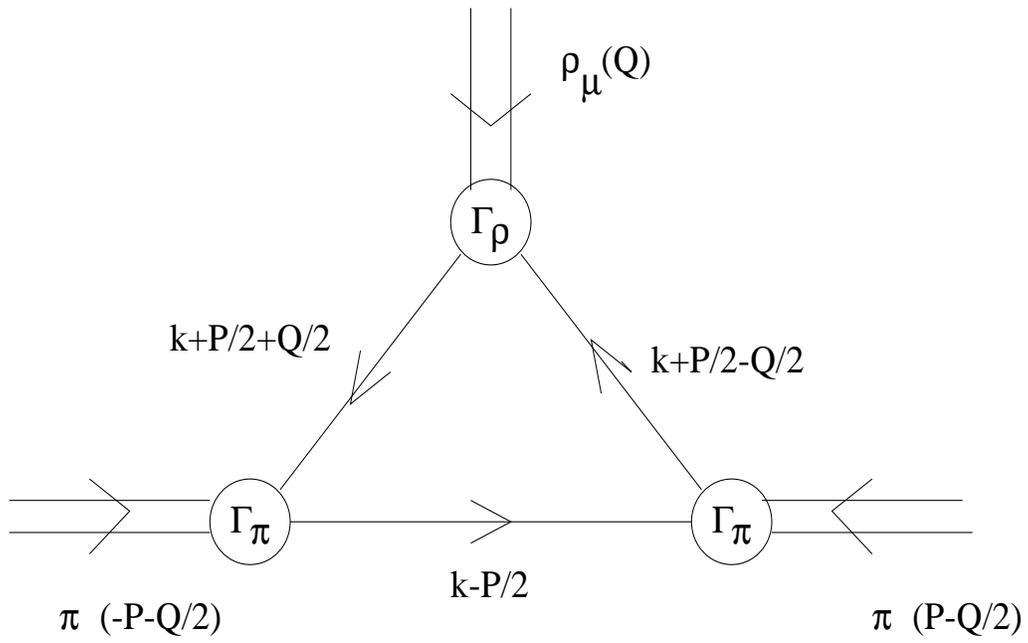,height=7.2in,width=7.2in}}
\caption{The quark loop mechanism for $\rho\pi\pi$ coupling.\label{rpp_tri}}
\end{figure}

\begin{figure}
\centerline{\psfig{figure=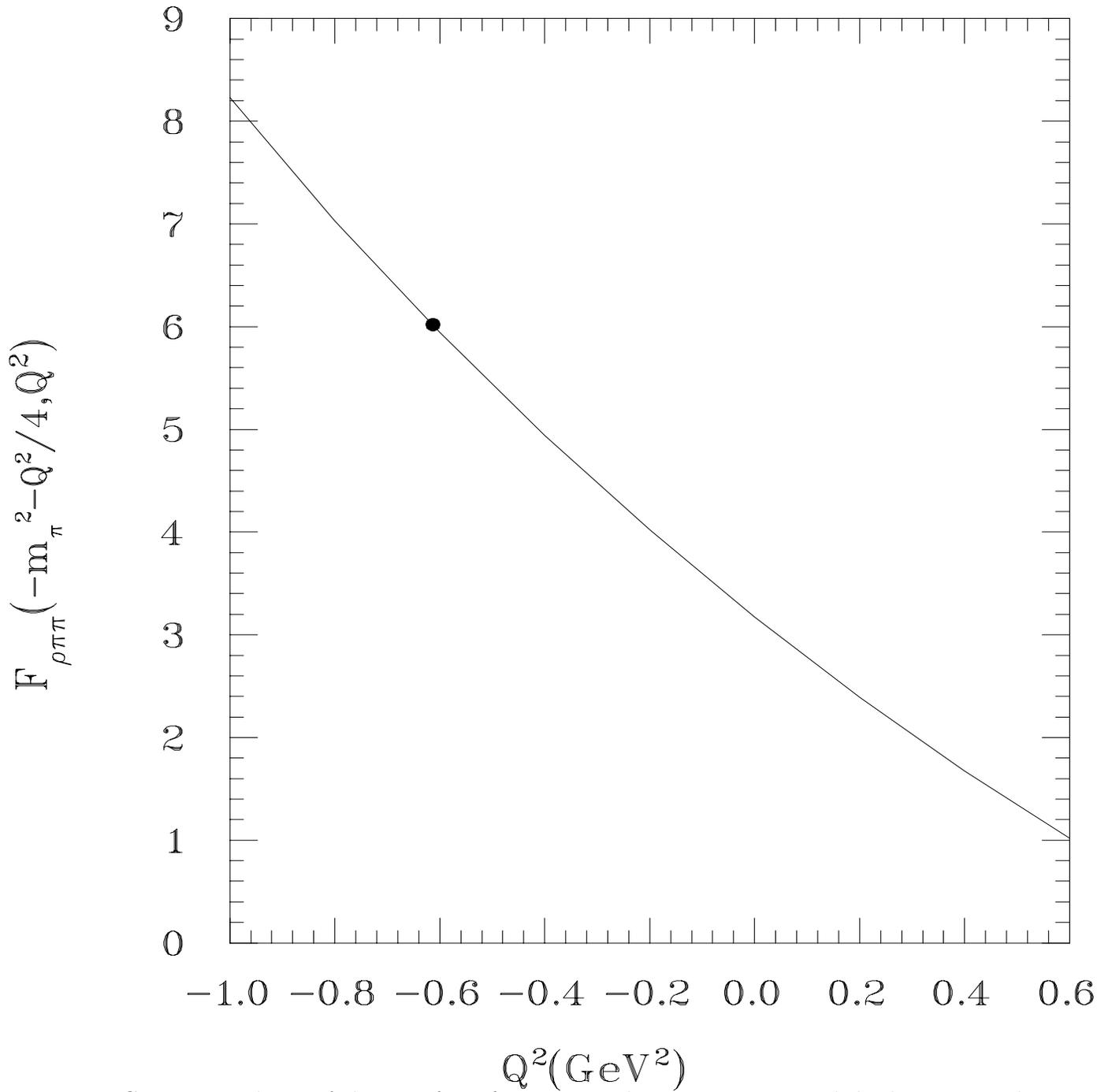,height=7.2in,width=7.2in}}
\caption{Dependence of the $\rho\pi\pi$  form factor upon the $\rho$
momentum with both pions on the mass-shell. The dot represents the
$\rho$ mass-shell.\label{rho}}  
\end{figure}

\begin{figure}
\centerline{\psfig{figure=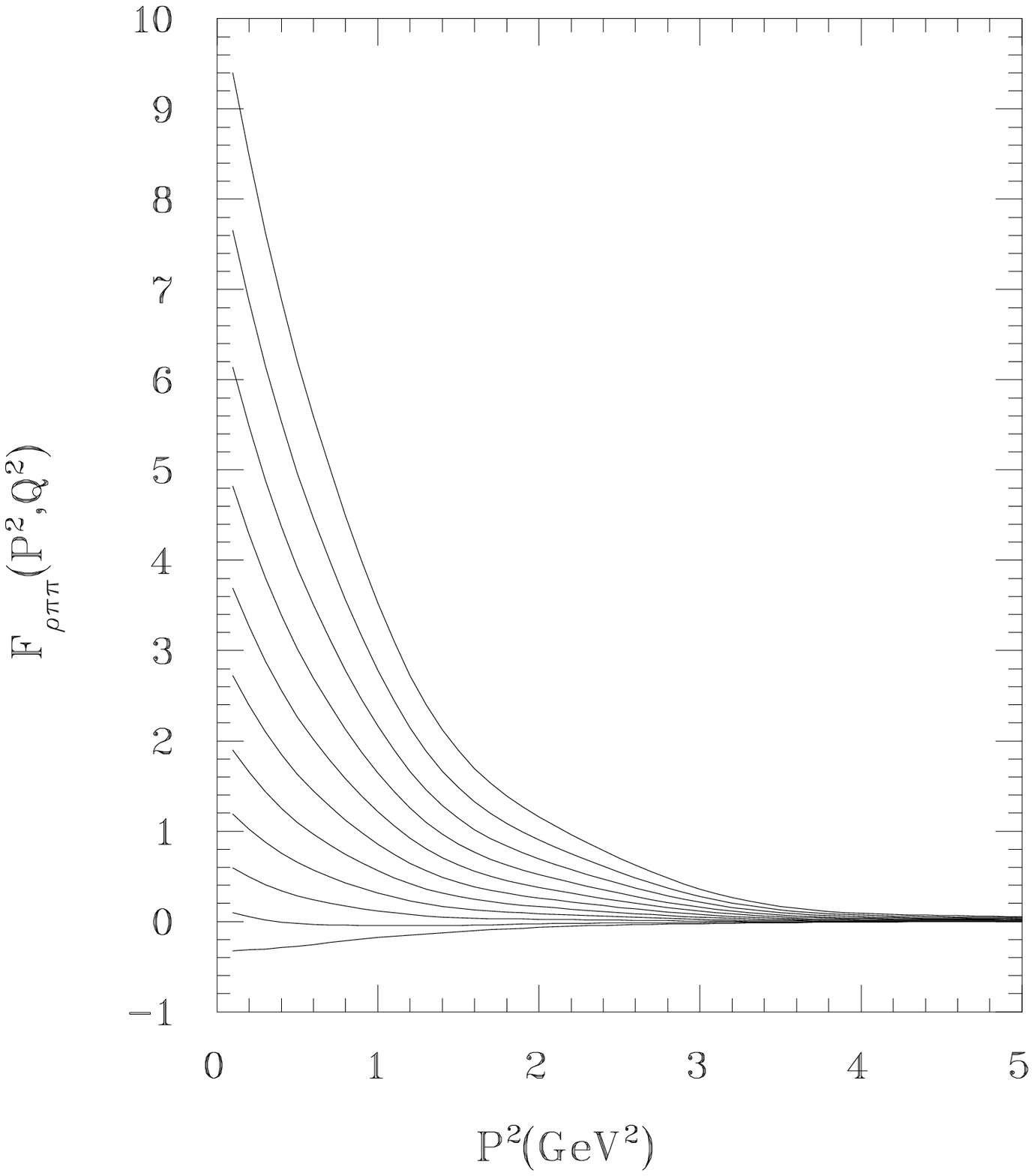,height=7.2in,width=7.2in}}
\caption{The $\rho\pi\pi$ form factor.  The top curve is for 
$Q^2=-1~{\rm GeV}^2$  and the bottom curve is for $Q^2=1~{\rm GeV}^2$.  Equal 
increments of $\Delta Q^2=0.2~{\rm GeV}^2$ are used in 
between.\label{rho_curves}}
\end{figure}

\begin{figure}
\centerline{\psfig{figure=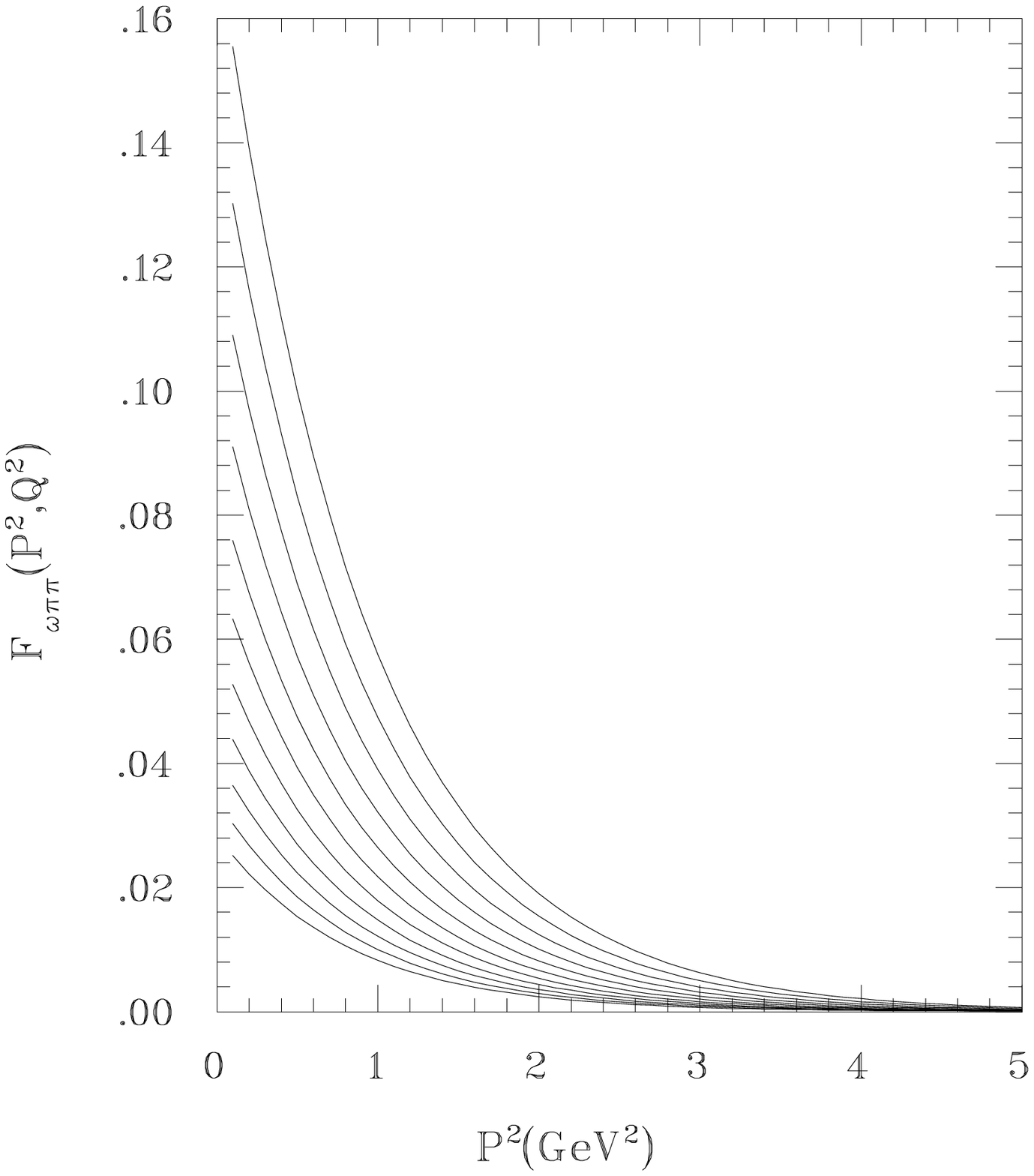,height=7.2in,width=7.2in}}
\caption{The $\omega\pi\pi$ form factor.  The top curve is for 
$Q^2=-1~{\rm GeV}^2$ and the bottom curve is for $Q^2=1~{\rm GeV}^2$.  Equal 
increments of $\Delta Q^2=0.2~{\rm GeV}^2$ are used in 
between.\label{omega_curves}}
\end{figure}

\begin{figure}
\centerline{\psfig{figure=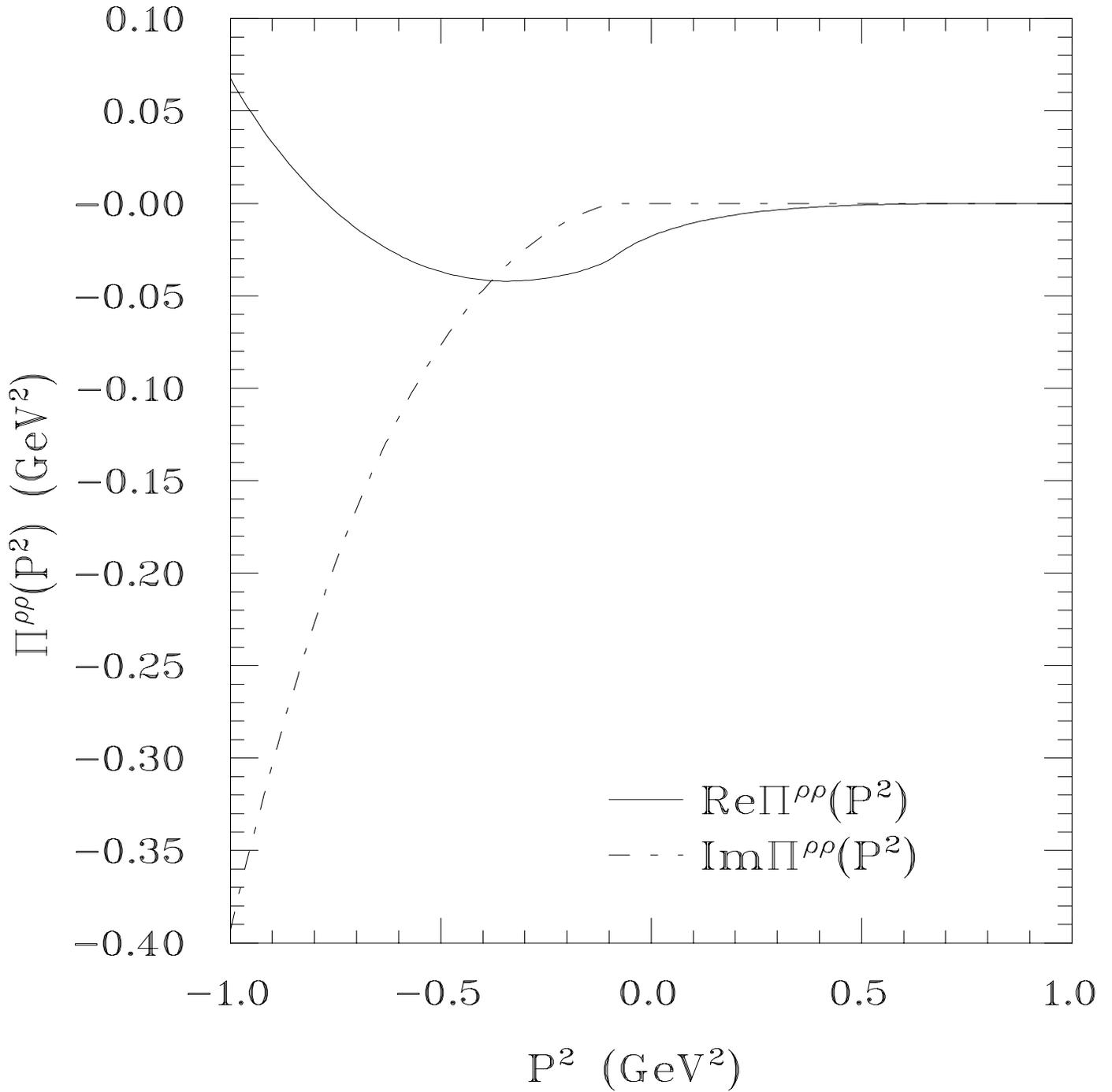,height=7.2in,width=7.2in}}
\caption{The real and imaginary parts of the pion loop contribution to the 
$\rho$ self-energy.\label{rho_pl}}
\end{figure}

\begin{figure}
\centerline{\psfig{figure=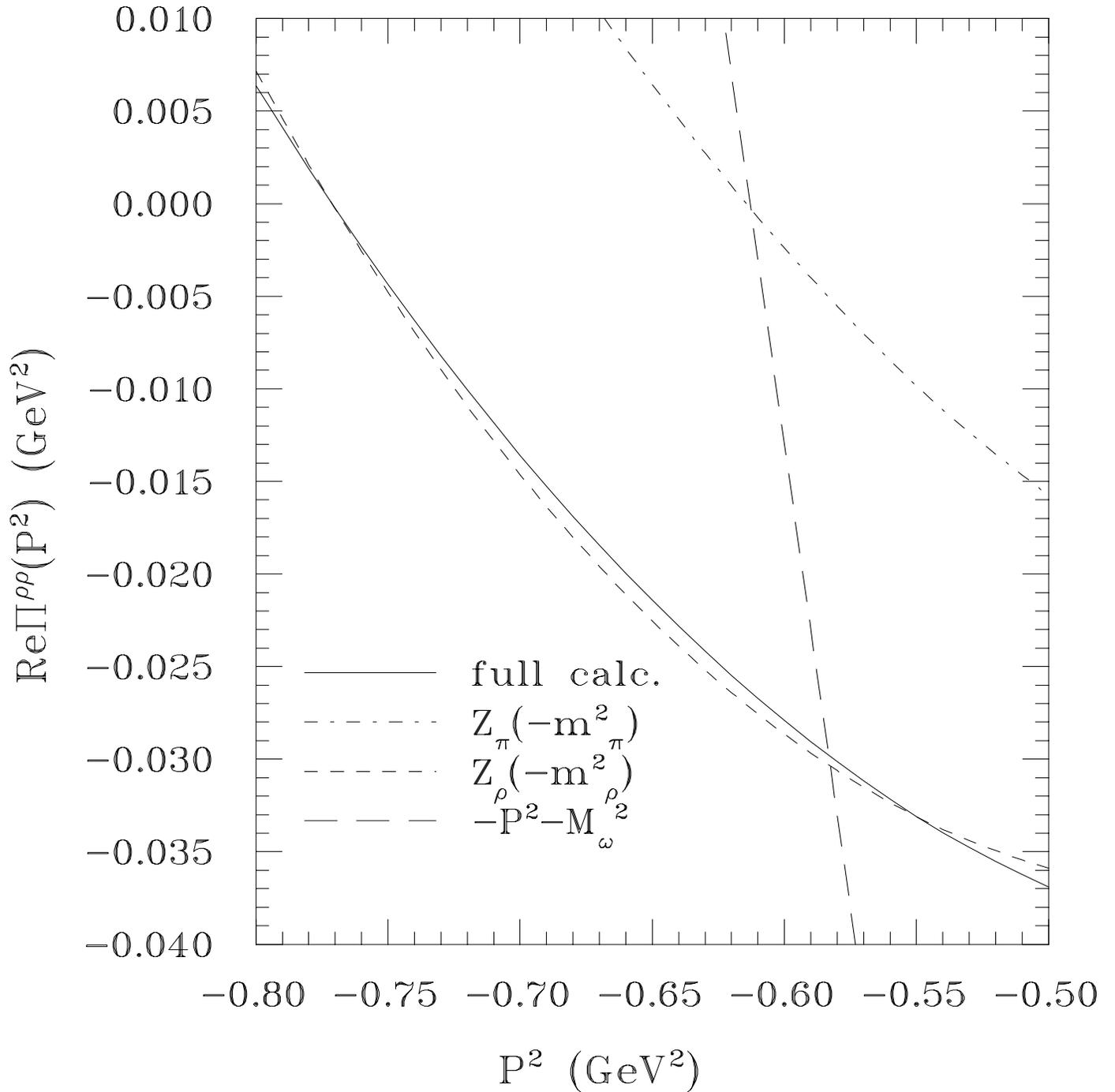,height=7.2in,width=7.2in}}
\caption{A graphical determination of the $\rho-\omega$ mass difference
and the influence of the $\bar{q}q$ composite nature of the $\pi$ and $\rho$
propagators as described in the text.\label{mcorr_comp}}
\end{figure}

\begin{figure}
\centerline{\psfig{figure=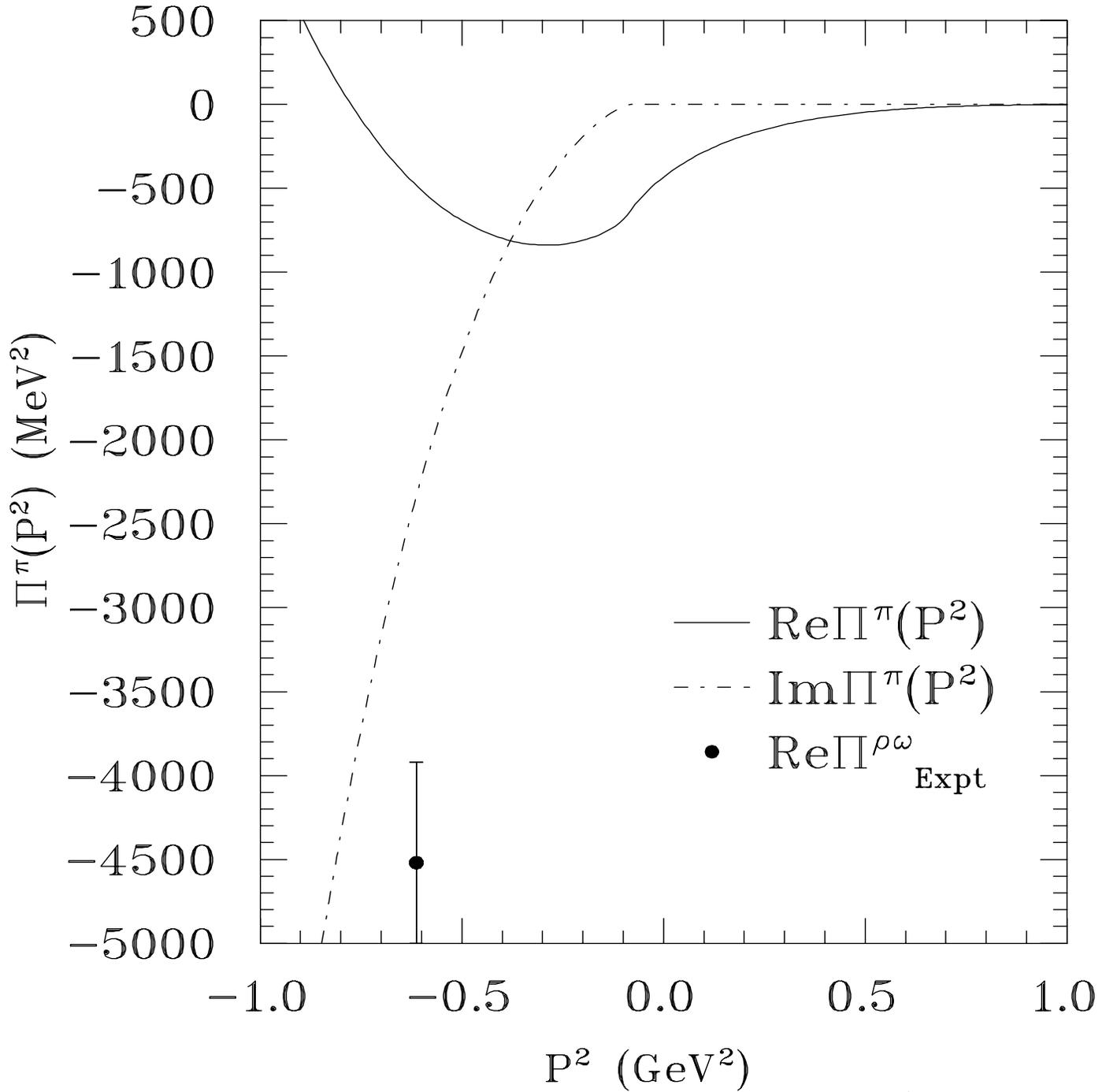,height=7.2in,width=7.2in}}
\caption{The real and imaginary parts of the pion loop contribution to the
$\rho^0-\omega$ mixing amplitude.\label{mix_pl}}
\end{figure}

\begin{figure}
\centerline{\psfig{figure=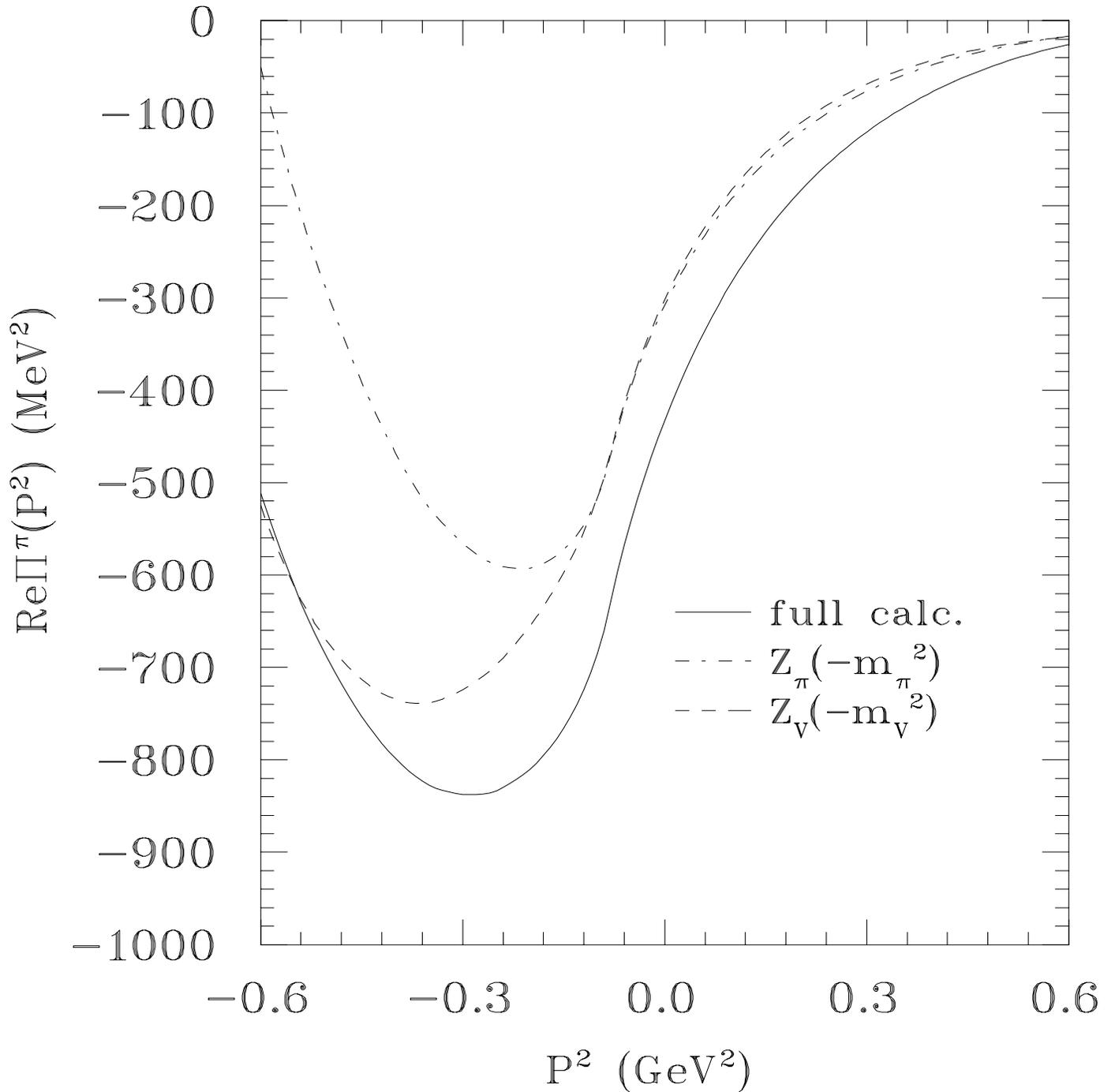,height=7.2in,width=7.2in}}
\caption{Various $\bar{q}q$ substructure effects on the pion loop contribution
to $\rho^0-\omega$ mixing as described in the text.\label{mix_comp}}
\end{figure}

\begin{figure}
\centerline{\psfig{figure=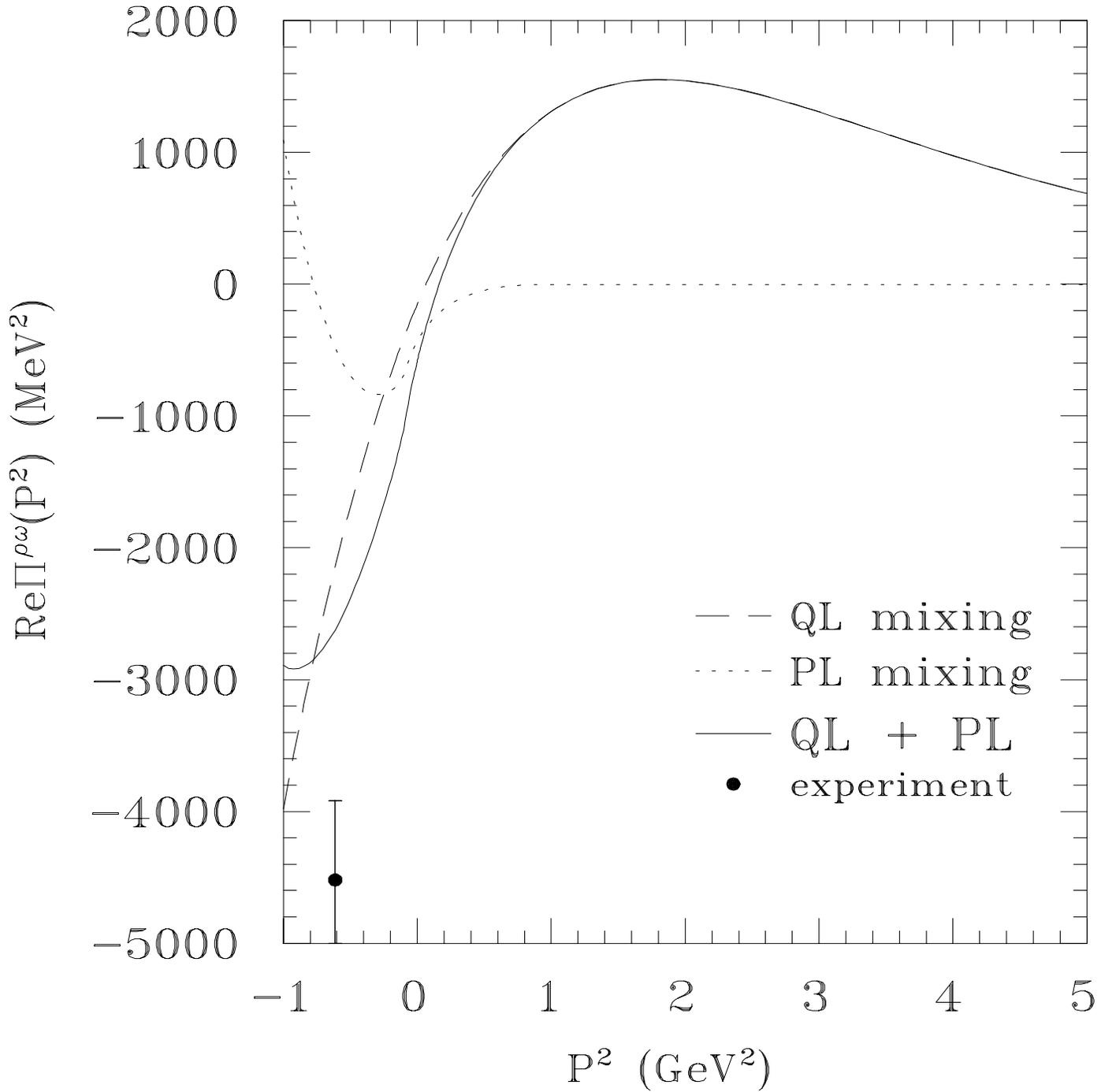,height=7.2in,width=7.2in}}
\caption{The quark loop (QL) and pion loop (PL) contributions to 
the $\rho^0-\omega$ mixing amplitude.\label{sum}} 
\end{figure}
\end{document}